\documentclass[longauth]{aa}

\usepackage{color}
\usepackage{graphicx}
\usepackage{natbib}
\usepackage{enumitem}
\usepackage{txfonts}
\usepackage{multirow}
\usepackage{rotating}
\usepackage{longtable}
\usepackage{amssymb}
\usepackage{verbatim}
\usepackage{caption}
\usepackage{subcaption}
\usepackage{marvosym}

\titlerunning{The role of environment in galaxies at 0.1$\leq$z$\leq$0.5.}
\authorrunning{V. Guglielmo et al.}
\begin{document}

\title{The XXL Survey: XXXVII. The role of the environment in shaping the stellar population properties of galaxies at 0.1$\leq$z$\leq$0.5.}

\author{V. Guglielmo \inst{1,2} \and B. M. Poggianti \inst{2} \and B. Vulcani \inst{2} \and S. Maurogordato \inst{3} \and J. Fritz \inst{4} \and M. Bolzonella \inst{5} \and S. Fotopoulou \inst{6} \and C. Adami \inst{7} \and M. Pierre \inst{8} 
} 

\institute{Max-Planck-Institut f{\"u}r Extraterrestriche Physik, Giessenbachstrasse, 85748 Garching, Germany\\
	\email{gglvnt@mpe.mpg.de}  
	\and INAF-Osservatorio Astronomico di Padova, Vicolo Osservatorio 5, 35122 Padova, Italy
    \and Observatoire de la C\^{o}te d’Azur, CNRS, Laboratoire Lagrange, Bd de l'Observatoire, Universit\'e C\^{o}te d’Azur, CS 34229, 06304 Nice cedex 4, France
    \and Instituto de Radioastronomia y Astrofisica, UNAM, Campus Morelia, A.P. 3-72, C.P. 58089, Mexico
    \and INAF, Osservatorio Astronomico di Bologna, Via Gobetti 93/3,
40129 Bologna, Italy
	\and Center for Extragalactic Astronomy, Department of Physics, Durham University,South Road,Durham DH1 3LE, UK
    \and Aix Marseille Universit\'e, CNRS, LAM (Laboratoire d'Astrophysique de Marseille) UMR 7326, F-13388, Marseille, France
    \and AIM,CEA,CNRS,Universit\'e Paris-Saclay, Universit\'e Paris Diderot, Sorbonne Paris Cit\'e, 91191 Gif-sur-Yvette, France
}

\date{Received xxx; accepted yyy}

\abstract
{
Exploiting a sample of galaxies drawn from the XXL-North multiwavelength survey, we present an analysis of the stellar population properties of galaxies at $0.1\leq z\leq 0.5$, by studying galaxy fractions and the star formation rate (SFR)-stellar mass (M$_\star$) relation. Furthermore, we exploit and compare two parametrisations of environment. When adopting a definition of ``global" environment, we consider separately cluster virial (r$\leq$1r$_{200}$) and outer (1r$_{200}<$r$\leq$3r$_{200}$) members and field galaxies. We also distinguish between galaxies that belong or do not belong to superclusters, but never find systematic differences between the two subgroups. When considering the ``local" environment, we take into account the projected number density of galaxies in a fixed aperture of 1 Mpc in the sky. 
We find that regardless of the environmental definition adopted, the fraction of blue or star-forming galaxies is the highest in the field or least dense regions and the lowest in the virial regions of clusters or highest densities. Furthermore, the fraction of star-forming galaxies is higher than the fraction of blue galaxies, regardless of the environment. This result is particularly evident in the virial cluster regions, most likely reflecting the different star formation histories of galaxies in different environments.
Also the overall SFR-M$_\star$ relation does not seem to depend on the parametrisation adopted. Nonetheless, the two definitions of environment lead to different results as far as the fraction of galaxies in transition between the star-forming main sequence and the quenched regime is concerned. In fact, using the {\it local} environment the fraction of galaxies below the main sequence is similar at low and high densities, whereas in clusters (and especially within the virial radii) a population with reduced SFR with respect to the field is observed.
Our results show that the two parametrisations adopted to describe the environment have different physical meanings, i.e. are intrinsically related to different physical processes acting on galaxy populations and are able to probe different physical scales.}

\maketitle

\section{Introduction}
\label{sec:introduction}

Observational studies aiming at understanding the processes that affect galaxy properties and determining the evolution of galaxies have been focussing more and more on the role played by both the environment in which a galaxy was formed  and that in which it is embedded for most of its lifetime \citep{Oemler1974,Dressler1980,Balogh2004b,Kauffmann2004,Baldry2006,Poggianti2009}.
In particular, galaxies that are gathered together and/or hosted in the potential well of dark matter haloes, together with those accreted from the cosmic web into bigger structures, undergo a variety of physical processes that may influence the timescale of star formation and stellar mass assembly. These processes are usually connected to the interaction between galaxies and the hot gas permeating the dark matter haloes of groups and clusters, or to galaxy-galaxy interactions \citep[e.g.,][and references therein]{Boselli2006,Boselli2014}.

One of the biggest challenges in observational studies aiming at describing the interplay between galaxies and their environment is the definition of the environment itself \citep{Haas2012,Muldrew2012,Etherington2015}.
Its parametrisation is commonly performed following two different strategies, which are able to probe different physical scales and have intrinsically different physical meanings.
The first approach is based on the potential well of dark matter haloes, and thus relies on physical properties of the cosmic structures such as the virial masses and radii, X-ray luminosity, and dynamical masses. According to this definition, which is commonly referred to as ``global" environment, going from the largest scale (i.e. the most massive haloes) in the cosmic web down to the scales of single galaxies we can define superclusters, clusters, groups, filaments, field, and voids.

The second description of environment is based on the computation of the projected over-density of galaxies and is referred to as ``local" environment.
Several methods have been explored for computing the local (projected) density of neighbouring galaxies, either based on computing the area enclosing the $Nth$ neighbour with respect to a central one or counting the number of galaxies enclosed within a fixed aperture. It has been shown that the latter methodology is closer to the real over-density measured in 3D space, more sensitive to high over-densities, less biased by the viewing angle, and more robust across cosmic times than the former \citep{Shattow2013}. For this reasons, we adopt this method to quantify the local environment.

Whatever the definition of environment, its strong connection with the observed properties of galaxies has been extensively demonstrated, both in terms of the average stellar age \citep[e.g.][]{Thomas2005, Smith2006} and the last episode of star formation \citep[and thus a lower fraction are continuing to form stars; e.g.][]{Lewis2002, Baldry2004, Balogh2004a, Balogh2004b, Kauffmann2004}.

Focussing on the intermediate redshift regime (0.25 $\leq$ z $\leq$ 1.2), colour fractions  have been found to depend strongly on the global environment; the incidence of blue galaxies is systematically higher in the field than in groups \citep{Iovino2010} and clusters \citep{Muzzin2012} and decreases with increasing absolute magnitude.
Similarly, also the mean star formation rate (SFR), specific-SFR (sSFR) and star-forming fraction are always higher in field galaxies than in clusters, decrease from the outskirts to the cluster  central region \citep{Treu2003,Poggianti2006,RaichoorAndreon2014,Haines2015} and depend on stellar mass in a given environment \citep{Muzzin2012}. 
Similar results have been found both in the local Universe \citep[e.g.][]{Balogh2004b} and at higher redshifts. Linking the star formation activity of galaxies with their cold molecular gas reservoir, \cite{Noble2017} discovered a population of massive cluster galaxies having higher gas fractions compared to the field, indicating a stronger evolution of massive haloes at high redshifts; a depletion of the cold gas reservoir emerges instead in a sample of z$\sim$0.4 cluster galaxies in \cite{Jablonka2013} with respect to field galaxies of the same stellar mass, with further decreasing trends towards the centre of the structures.

Considering instead the local density (LD) parametrisation, the colour and star-forming fractions have also found to be lower in denser environments, both in the local Universe \citep[e.g,][]{Balogh2004a, Baldry2006} and at intermediate redshifts \citep[e.g.,][]{Cooper2008,Cucciati2006,Cucciati2010,Cucciati2017}.
However, \cite{Darvish2016} found that in the star-forming population the median SFR and sSFR are similar  at different values of the local density, regardless of redshift and galaxy stellar mass up to z$\sim$3, and \cite{Elbaz2007} even advocated the increase of the SFR of galaxies at z$\sim$1 in denser environments.

The effect of global or local environment on galaxy properties has also been investigated in terms of the relation between the SFR and galaxy stellar mass.
The existence of a tight relation of direct proportionality between SFR and galaxy stellar mass (SFR–M$_\star$) and sSFR–M$_\star$ has been established from z=0 out to z>2, with a roughly constant scatter of $\sim$0.3 dex out to z$\sim$1 \citep{Brinchmann2004,Daddi2007,Noeske2007,Salim2007,Rodighiero2011,Whitaker2012,Sobral2014,Speagle2014}. Star-forming galaxies lie on the so-called main sequence, whereas the quenched population occupy a locus with little or non-detectable SFR.

The representation of the SFR-M$_\star$ plane is necessary to understand the characteristics of the star-forming population of galaxies in different environments and to analyse whether the process leading to the shutting down of the star formation activity in a galaxy (and thus its transformation into a passive galaxy) proceeds similarly in different environments and whether the definition of the environment itself plays a role. 
In fact, fast quenching processes would leave the cluster/high-density regions SFR-M$_\star$ relation unperturbed with respect to the field/low-density regions, leaving the median SFR in agreement at all stellar masses. In contrast, slow quenching mechanisms would increase the number of galaxies with reduced SFRs shifting the overall distribution of SFRs towards lower values than those of main sequence galaxies of similar mass. 

When inspecting the SFR-M$_\star$ relation in different global environments, 
a population of low star-forming galaxies in a transition stage between the main sequence and the quenched population (hereafter ``transition" galaxies) has been observed in clusters at all redshfits up to z<0.8 \citep{Patel2009, Vulcani2010, Paccagnella2016}. This population is missing in the field.
In particular, \cite{Paccagnella2016} found that at 0.04$<z<$0.07 galaxies in transition are preferentially found within the virial radius ($R_{200}$), and their incidence increases at distances $<0.6R_{200}$. These galaxies are older and present redder colours than galaxies in the main sequence and show reduced mean SFRs over the last 2-5 Gyr, regardless of their stellar mass.
Moreover, using spatially resolved observations from SDSS-IV MaNGA, \cite{Belfiore2017} associated the transition population with a population of galaxies having central low ionisation emission-line regions, resulting from photoionisation by hot evolved stars, and star-forming outskirts. These galaxies are preferentially located in denser environments such as galaxy groups and are undergoing an inside-out quenching process.

On the contrary, studies on galaxy samples based on a local parametrisation of environment do not find differences in the SFR-M$_\star$ of galaxies at different densities (\citealt{Peng2010,Wijesinghe2012}, but see \citealt{Popesso2011} at high z).

It is important to stress however that  different results in the literature obtained by adopting different parametrisations of the environment are hard to compare, either because of the different selection criteria on the samples or custom definitions used to define, for example, the local galaxy over-density.

The aim of this work is to study the star formation properties and colours of galaxies adopting different definitions of environment, to acquire a general understanding of the phenomena that characterise and influence the observed properties of galaxies at different epochs and in different conditions.
The main questions we want to address are: 1) How do the star-forming and blue fractions depend on environment? 2) Are there differences in the star-forming population in different environments? Namely, are star-forming galaxies in clusters or dense environments as star-forming as galaxies in the field or lower density environments? 3) How does the definition of the environment itself affects these tracers?

We characterise galaxies in three redshift bins from z=0.1 up to z=0.5, in X-ray massive groups and clusters ($1.13\times 10^{13} \leq M_{200}/M_\odot \footnote{$M_{200}$ is the mass of a virialised structure, i.e. the mass budget inside the virial radius, which corresponds to that radius within which the material is virialised and external to which the mass is still collapsing onto the object. Some simulations suggest that this occurs at a density contrast of 200 with respect to the critical density of the Universe $\rho_c$, more or less independently of cosmology.} \leq 9.28 \times 10^{14}$, hereafter simply clusters) observed in the XXL Survey.  
This survey \citep[hereafter XXL Paper I]{Pierre2016}, is an extension of the XMM-LSS 11 $\rm deg^2$ survey \citep{Pierre2004}, consisting of 622 XMM pointings covering two extragalactic regions of $\sim 25 \, {\rm deg^2}$ each, one equatorial (XXL-N) and one in the southern hemisphere (XXL-S). The survey reaches a sensitivity of $\rm \sim 6 \times 10^{-15} erg \, s^{-1} \, cm^{-2}$ in the [0.5-2] keV band for point sources.

This study is focussed on computing the fraction of star-forming and blue galaxies and the SFR-M$_\star$ relation, in the field versus clusters, also distinguishing between structures belonging or not to superclusters, and as a function of LD.
The paper is organised as follows: in Section \ref{sec:data_samples} we present the catalogues of clusters and galaxies, the tools used to compute galaxy stellar population properties and the computation of the spectroscopic incompleteness weights; in Section \ref{sec:pops} we characterise different galaxy populations on the basis of their SFR and colours; in Section \ref{sec:results_global} we explore the dependence of the stellar population properties on global environment, performing a detailed analysis on galaxy fractions (Sect. \ref{sec:SFing_blue_frac}) and on the SFR-M$_\star$ relation (Sect. \ref{sec:sfr_m_glob} and \ref{sec:F_TR_global}); in Section \ref{sec:results_LD} we analyse the galaxy population properties as a function of local environment, following the same scheme as Sect. \ref{sec:results_global}.
In section \ref{sec:discussion} we discuss our results obtained with the two parametrisations of environments regarding the galaxies in transitions (Sect. \ref{sec:discussion_F_tr}) and the ratio of star-forming to blue fractions (Sect. \ref{sec:Discussion_SFing_vs_blue}). Finally, we  present our conclusions in Sect. \ref{sec:conclusions}.

Throughout the paper we assume $\rm H_0 = 69.3 \, km\, s^{-1}\, Mpc^{-1},\, \Omega_{m} =0.29,\, \Omega_{\Lambda} =0.71$ \citep[][Planck13+Alens]{Planck2014}. We adopt a \cite{Chabrier2003} initial mass function (IMF) in the mass range $0.1-100 M_\odot$.

\section{Data samples and tools}
\label{sec:data_samples}

\subsection{Catalogue of structures}

Our environmental study is grounded in X-ray selected clusters from the XXL survey (XXL Paper I).
The selection of the cluster candidates starting from X-ray images was presented by \cite{Pacaud2016} (hereafter XXL Paper II).

By means of the \textsc{Xamin} pipeline \citep{Pacaud2006}, each structure is assigned to a specific detection class on the basis of the level of contamination from point sources. 
Class 1 (C1) clusters are the highest surface brightness extended sources, which have no contamination from point sources; Class 2 (C2) clusters are extended sources that are fainter than those classified as C1 and have a 50\% contamination rate before visual inspection. Contaminating sources include saturated point sources, unresolved pairs, and sources strongly masked by CCD gaps, for which not enough photons were available to permit reliable source characterisation. Class 3 (C3) are (optical) clusters associated with an X-ray emission that is too weak to be characterised, and whose selection function is therefore undefined. 

The spectroscopic confirmation and redshift assignment  of cluster candidates are presented in \cite{Adami2018} (hereafter XXL Paper XX, but see also \citealt{Guglielmo2018a}, hereafter XXL Paper XXII). The procedure is similar to that already used for the XMM-LSS survey (e.g., \citealt{Adami2011}), and is based on an iterative semi-automatic process. 
The final catalogue of spectroscopically confirmed extended sources contains 365 clusters, 207 ($\sim 56\%$) of which are classified as C1, 119 ($\sim 32\%$) as C2 and the remaining 39 ($\sim 11\%$) are C3. For the reasons explained above, C3 clusters are not included in the current work. 
A larger subsample of objects with respect to the first data release \citep[][XXL Paper III]{Giles2016} underwent a direct X-ray spectral measurement of luminosity and temperature, down to a lowest flux of $\sim$2 $\times$10$^{-15}$  $\rm erg \, s^{-1} \, cm^{-2}$ in the [0.5-2] keV band and within 60 arcsec (235 clusters).

To have homogeneous estimates for the complete sample, and as already performed in \cite{Guglielmo2018a,Guglielmo2018b} (hereafter XXL Paper XXX), we used the cluster properties derived through scaling relations\footnote{All the cluster quantities derived through scaling relations are therefore named using the suffix ``scal".} starting from the X-ray count-rates.
The method is presented in XXL Paper XX, from which (Table F.1) we extracted the values of the X-ray temperature ($T_{300kpc,scal}$), $r_{500,scal}$\footnote{$r_{500,scal}$ is defined as the radius of the sphere inside which the mean density is 500 times the critical density $\rho_c$ of the Universe at the cluster redshift.}, $M_{500,scal}$\footnote{$M_{500,scal}$ derives from $r_{500,scal}$ and is defined as $4/3 \pi 500 \rho_c r_{500,scal}^3$}. 
The luminosity in the 0.5-2.0 keV range ($L^{XXL}_{500,scal}$) was not published in Paper XX but is available internally to our collaboration.
XXL Paper XXII derived the virial mass $M_{200}$ from $M_{500,scal}$ using the recipe given in \cite{Balogh2006}, and computed the velocity dispersion ($\sigma_{200}$) through the relation given in \cite{Poggianti2006}, based on the virial theorem.

In XXL Paper XX, 35 superclusters were identified in both XXL-N and XXL-S fields in the 0.03$\leq$z$\leq$1.0 redshift range, by means of a friend-of-friend (FoF) algorithm characterised by a Voronoi tesselation technique.
The physical associations with at least three clusters are called ``superclusters''. All the details of the methodology are provided in XXL Paper XX.

In this work we focus on clusters observed in the XXL-N region at 0.1$\leq$z$\leq$0.5. The sample is composed of 111 clusters that are fully characterised in terms of X-ray luminosities, temperatures, virial masses, and radii. Of these structures, 68 ($\sim 60\%$) belong to superclusters, thus it is possible to study the impact of the large-scale structure on galaxy properties. To do so, we treat separately galaxies that belong or do not belong to a supercluster, and call these ``(S)'' and ``(NS)'', respectively.
Taking as a reference the nomenclature adopted in XXL Paper XX, the superclusters considered in this work are reported in Table \ref{tab:Superclusters}.

\begin{table*}
\begin{center}
\caption{List of superclusters detected in XXL Paper XX and included in our sample. The first column is the name of the supercluster according to XXL Paper XX nomenclature, the second and third columns are the centroid coordinates (J2000.0 equinox) the fourth column is the mean redshift, and the last column is the list of clusters belonging to each supercluster.}
\begin{tabular}{ccccr}
\hline
Name & RA & DEC & z$_{mean}$ & Members \\
& (deg) & (deg) & & (XLSSC number) \\
\hline
XLSSsC N01 & 36.954 & -4.778 & 0.296 & 008,013,022,024,027,028,070,088,104,140,148,149,150,168 \\
XLSSsC N02 & 32.059 & -6.653 & 0.430 & 082,083,084,085,086,092,093,107,155,172,197 \\
XLSSsC N03 & 32.921 & -4.879 & 0.139 & 060,095,112,118,138,162,176,201 \\
XLSSsC N06 & 33.148 & -5.568 & 0.300 & 098,111,117,161,167 \\
XLSSsC N07 & 36.446 & -5.142 & 0.496 & 020,049,053,143,169 \\
XLSSsC N08 & 36.910 & -4.158 & 0.141 & 041,050,087,090 \\
XLSSsC N09 & 37.392 & -5.227 & 0.190 & 074,091,123,151 \\
XLSSsC N10 & 36.290 & -3.411 & 0.329 & 009,010,023,129 \\
XLSSsC N11 & 34.438 & -4.867 & 0.340 & 058,086,192 \\
XLSSsC N12 & 34.138 & -5.003 & 0.447 & 110,142,144,187 \\
XLSSsC N15 & 34.466 & -4.608 & 0.291 & 126,137,180,202 \\
XLSSsC N16 & 36.156 & -3.455 & 0.174 & 035,043,182 \\
XLSSsC N17 & 34.770 & -4.240 & 0.203 & 077,189,193 \\
XLSSsC N18 & 30.430 & -6.880 & 0.336 & 156,199,200 \\
XLSSsC N19 & 35.629 & -5.146 & 0.380 & 017,067,132 \\
\hline
\end{tabular}
\label{tab:Superclusters}
\end{center}
\end{table*}

\begin{figure}
\centering
\includegraphics[scale=0.37]{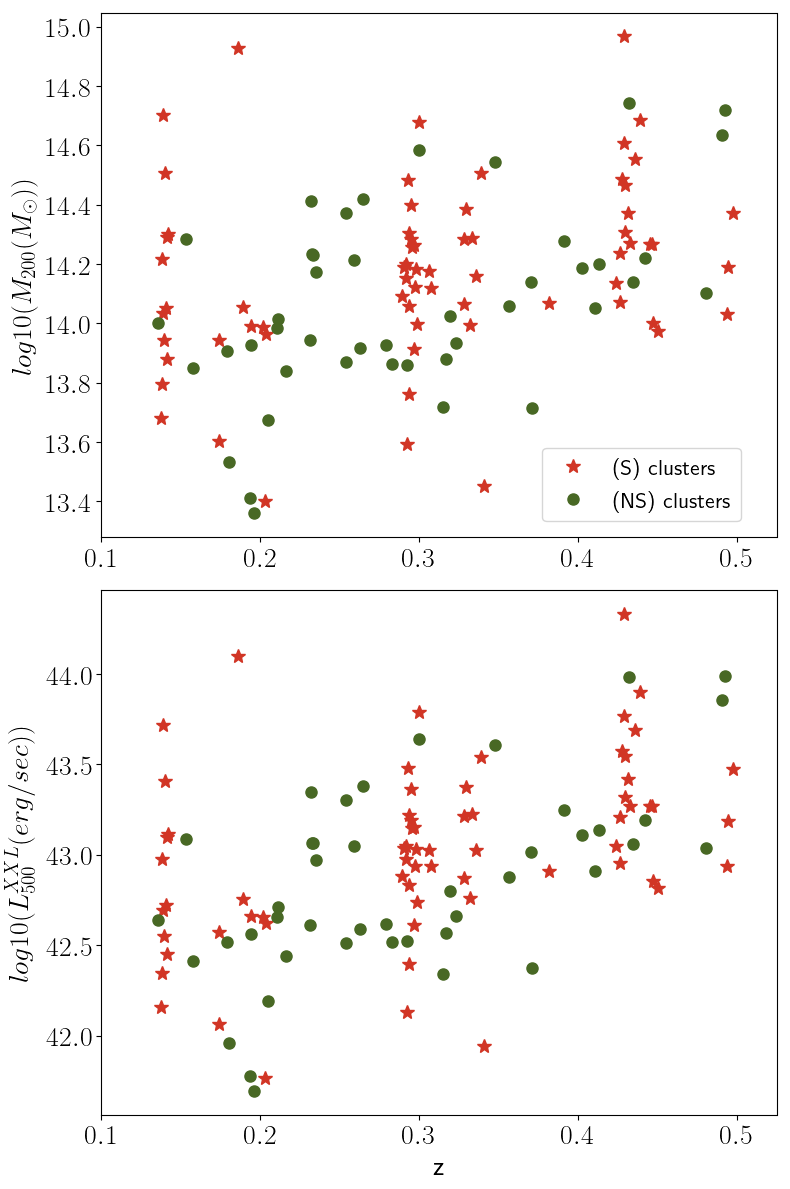}
\caption{$M_{200}$ (top), $L^{XXL}_{500}$ (bottom) versus redshift for the 111 XXL-N C1+C2  clusters at 0.1$\leq$z$\leq$0.5. Clusters that belong to superclusters are represented by red stars, cluster that do not belong to any superclusters are represented by green points.}
\label{cluster_fig}
\end{figure}

Figure \ref{cluster_fig} shows how M$_{200}$ and L$_{500}^{XXL}$ vary with redshift within the sample, for clusters within and outside superclusters. As already mentioned in XXL Paper XXII, selection effects emerge: at $z>0.4$ the survey detects only the most massive clusters ($M_{200}\geq 10^{14} M_\odot$). Nonetheless, no systematic differences are detected between (S) and (NS) clusters.

\subsection{Galaxy catalogue} 
\label{gal_cat_w1}

We made use of the galaxy properties included in the spectrophotometric catalogue presented in XXL Paper XXII. As for the catalogue of structures, we focussed on the XXL-N region and on the redshift range 0.1$\leq$z$\leq$0.5.

The photometric and photo-z information in XXL-N were mainly taken from the CFHTLS-T0007 photo-z catalogue in the W1 Field ($\rm 8^\circ \times 9^\circ$, centred at RA=34.5000$^\circ$ and DEC=-07.0000$^\circ$). The data cover the wavelength range 3500\AA $< \lambda < 9400$\AA $\,$ in the $u^*$, $g^\prime$, $r^\prime$, $i^\prime$, and $z^\prime$ filters. Photometric data for a number of galaxies in the spectroscopic database that did not have any correspondence in the CFHTLS catalogue were taken from \citet{Fotopoulou2016}. This catalogue contains aperture magnitudes in the $g^\prime$, $r^\prime$, $i^\prime$, $z^\prime$, $J^\prime$, $H^\prime$, and $K^\prime$ bands that have been converted into total magnitudes using a common subsample of galaxies with the CFHTLS-T0007 W1 field catalogue (see XXL Paper XXII).

All magnitudes are Sextractor \verb!MAG_AUTO! magnitudes \citep{Bertin1996} in the AB system corrected for Milky Way extinction according to \cite{Schlegel1998}.
The error associated with photo-z in the magnitude range we are probing in this work (r$<$20.0, see XXL Paper XXII and below) is redshift dependent, and according to the CFHTLS-T0007 data release document, is $\sigma/(1+z) \sim 0.031$.

Spectroscopic redshifts are hosted in the XXL spectroscopic database that is included in the CeSAM (Centre de donn\'eeS Astrophysiques de Marseille) database in Marseille.\footnote{http://www.lam.fr/cesam/} 
As described in XXL Paper XXII, the database collects spectra and redshifts coming from different surveys covering the XXL pattern (mainly GAMA, SDSS, VIPERS, VVDS, VUDS, and XXL dedicated spectroscopic campaigns, see Table 2 in XXL Paper XXII), and the final spectroscopic catalogue was obtained by removing duplicates using a careful combination of selection criteria (the so-called priorities) and  accounting for the quality of the spectra (i.e. the parent survey) and  of the redshift measurement.
Overall, the uncertainties on the galaxy redshift in the database vary from 0.00025 to 0.0005, as computed from multiple observations of the same object; we  consider the highest value in this range as the typical redshift error for all objects.
We note that the spectroscopic catalogue did not undergo any preselection or flag assignment to identify active galactic nuclei (AGN), and thus our sample may be contaminated by the presence of such peculiar sources. We address this point in more detail and quantify the contribution of AGNs later in this paper.

\begin{figure*}
\centering
\includegraphics[scale=0.65]{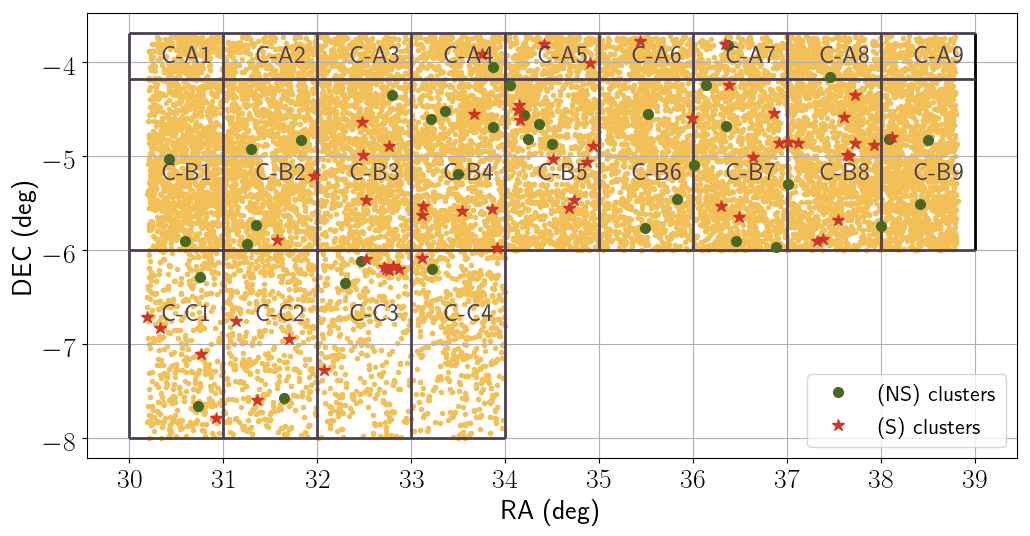}\caption{Spatial distribution in the XXL-N area of  galaxies in the spectrophotometric sample (yellow dots) and of X-ray confirmed clusters. The clusters in superclusters are reprensented with red stars and the clusters outside superclusters with green points.  The region is divided into 22 cells (named as indicated inside each cell), used to compute the spectroscopic completeness (see details in Appendix \ref{compl_w1}).}
\label{RA_DEC_sample}
\end{figure*}

The final galaxy sample is obtained from the crossmatch between the photometric and spectroscopic sample. Figure \ref{RA_DEC_sample} shows the distribution of galaxies and clusters  in the coordinates plane, for the magnitude limited sample that is presented below.

\subsection{Tools}

The stellar population properties of galaxies were derived relying on either their photometric or spectroscopic data.
In the first case, we made use of the spectral energy distribution (SED) fitting code LePhare\footnote{http://www.cfht.hawaii.edu/~arnouts/lephare.html} \citep{Arnouts1999,Ilbert2006} to compute absolute magnitudes, and therefore rest-frame colours, as described in XXL Paper XXII.
In the second case, we fit galaxy spectra via \textsc{SINOPSIS}\footnote{http://www.crya.unam.mx/gente/j.fritz/JFhp/SINOPSIS.html} (SImulatiNg OPtical Spectra wIth Stellar population models), a spectrophotometric fitting code fully described in \citet{Fritz2007, Fritz2011,Fritz2017} and already largely used to derive physical properties of galaxies in many samples \citep{Dressler2009,Vulcani2015,Guglielmo2015,Paccagnella2016,Paccagnella2017,Poggianti2017}. Among the outputs of the model, we considered SFRs and galaxy stellar masses ($M_*$), defined as the mass locked into stars, both those which are still in the nuclear-burning phase, and remnants such as white dwarfs, neutron stars, and stellar black holes.

While LePhare could be applied to the whole spectrophotometric sample of galaxies (provided that the catalogue contains magnitudes at least in two filters for each objects), \textsc{SINOPSIS} was run on the subsample of galaxies that have either SDSS or GAMA spectra, which are flux calibrated and have the best available spectral quality.
As discussed in \cite{Fritz2014}, in the lowest resolution spectra of this work, i.e. GAMA spectra, emission lines can be measured down to a limit of 2 \AA, while any emission measurement below this threshold is considered unreliable. In terms of sSFR, this sets a lower limit of $10^{-12.5} yr^{-1}$.

The final sample is composed of
galaxies with reliable outputs coming from  both LePhare and SINOPSIS.

\begin{table}
\centering
\caption{Number of galaxies above the magnitude and mass completeness limits in three redshift bins. The quantities in parentheses refer to the number of galaxies weighted for spectroscopic completeness. Values of $M_{lim}$ are given in the main text.\label{tab:sampleall}}
\begin{tabular}{ccc}
\hline
 zbin &  $r \leq 20$ & $\rm M_\ast>M_{lim}$ \\
 \hline
$0.1 \leq z<0.2$ & 6132 (11426) & 5438 (10117) \\
$0.2 \leq z<0.3$ & 5438 (11601) & 7490 (7803) \\
$0.3 \leq z\leq0.5$ &  4777 (7902) & 3352 (5593) \\
\hline 
all & 18399 (30929) & 13857 (21303) \\
\end{tabular}
\end{table}

\begin{figure*}
\begin{center}
\includegraphics[scale=0.69]{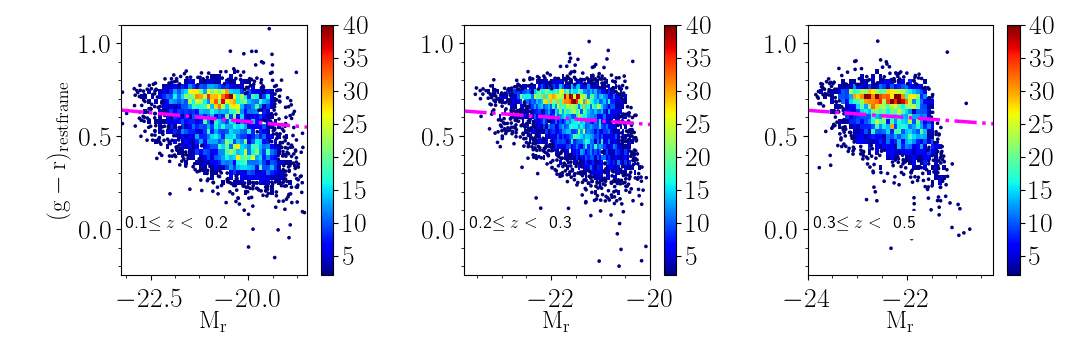}
\caption{Colour-magnitude diagrams in the magnitude limited sample in the three redshift bins analysed, with increasing redshift from left to right as indicated in the labels. Single galaxies are plotted as blue dots, while galaxies in higher density regions are grouped together and plotted as rectangles colour-coded according to their number density as indicated in the colour bar located on the side of each panel. The magenta dotted line shows the separation between red and blue objects using the (g-r)$_{rest-frame}$ colour.}
\label{CMD_w1}
\end{center}
\end{figure*}

\subsection{Samples and spectroscopic completeness}

In what follows, we consider galaxies in three redshift bins, $0.1\leq z < 0.2$, $0.2\leq z <0.3$, $0.3\leq z \leq 0.5$ and study both magnitude and mass limited samples. As detailed in XXL Paper XXII, magnitude completeness limit was set to an observed magnitude of $r=20.0$ at all redshifts, and is converted into a different mass completeness limit at each redshift. To determine this limit, at each redshift we converted the observed magnitude limit into a rest-frame magnitude limit and computed the mass of an ideal object having the faintest magnitude and the reddest colour in that redshift bin. Following XXL Paper XXII, the stellar mass limit of each redshift bin is that corresponding to the lowest limit of each interval; i.e. at 0.1$\leq$z$<$0.2 is the stellar mass limit corresponding to z=0.1.
We therefore adopted the following values:
\noindent
\begin{itemize}
\item[]- 0.1 $\leq$ z $<$ 0.2: $\rm M_\star>10^{9.5} M_\odot$ 
\item[]- 0.2 $\leq$ z $<$ 0.3: $\rm M_\star>10^{10.3} M_\odot$ 
\item[]- 0.3 $\leq$ z $\leq$ 0.5: $\rm M_\star>10^{10.8} M_\odot$ 
\end{itemize}

The galaxy magnitude complete sample  includes 18399 galaxies, the mass complete sample includes 13857 galaxies. Table \ref{tab:sampleall} reports the number of galaxies in the different redshift bins for both samples.
Both raw numbers and those corrected for incompleteness are given. The method used to compute the spectroscopic completeness is described in Appendix \ref{compl_w1}. Briefly, as the spectroscopic sample spans a relatively wide redshift range, we sliced the sample into different redshift bins and  quantified the number of galaxies that fall/are expected to fall into that given redshift bin, based on both spectroscopic and photometric redshifts.
As already performed in XXL Paper XXII, we accounted for the change in the spectroscopic sampling of different surveys by dividing the sky into 22 cells (shown in Fig. \ref{RA_DEC_sample}), and in intervals of 0.5 r-band magnitude within each cell.
The completeness curves resulting from this computation were converted into completeness weights which are attributed to each galaxy given its redshift, astrometry, and magnitude.

\section{Galaxy subpopulations} \label{sec:pops}

In our analysis we characterised separately the star-forming properties and rest-frame colours of galaxies in different environments and at different redshifts. We therefore need to define two different criteria to separate star-forming/blue galaxies from passive/red galaxies. 

First, we considered as ``star forming'' those galaxies with $\rm sSFR=SFR/M_\star > 10^{-12} yr^{-1}$ and ``passive'' the remaining galaxies. We point out that this sSFR threshold is the same in the three redshift bins considered, which is justified by the scarce evolution in the sSFR-stellar mass plane in this redshift range \citep[see e.g.][]{Whitaker2012}.

Then, we considered as ``blue'' galaxies those whose rest-frame colour is bluer than a certain threshold, and ``red'' the rest. 
To identify such threshold in colour, we investigated the relation between the $\rm (g-r)_{rest-frame}$ colour and absolute magnitude $\rm M_r$, in the three redshift bins separately.
Figure \ref{CMD_w1} shows the rest-frame colour-magnitude diagram (CMD) in each redshift bin.
\begin{figure}
\begin{center}
\includegraphics[scale=0.55]{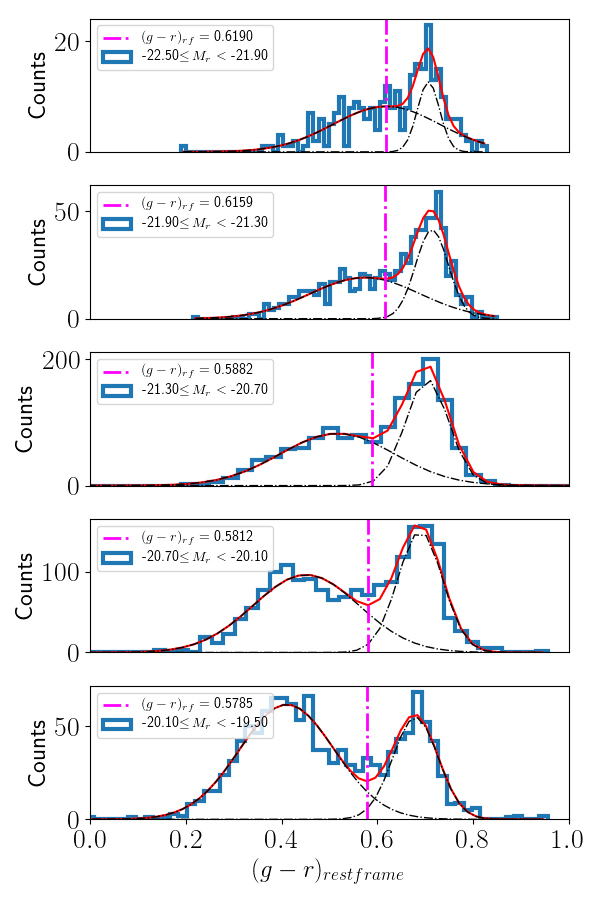}
\caption{Rest-frame (g-r) colour distributions in five absolute magnitude bins for galaxies at $0.1 \leq z < 0.2$. The red curve shows the double-Gaussian fit performed on the distributions and the single Gaussians are represented with the black dashed line. The magenta vertical lines indicate the local minima in the valley between the two Gaussian peaks, and define the separation between  the red sequence and blue cloud.}
\label{CMD_fit_0102} 
\end{center}
\end{figure}
\begin{figure}
\begin{center}
\includegraphics[scale=0.5]{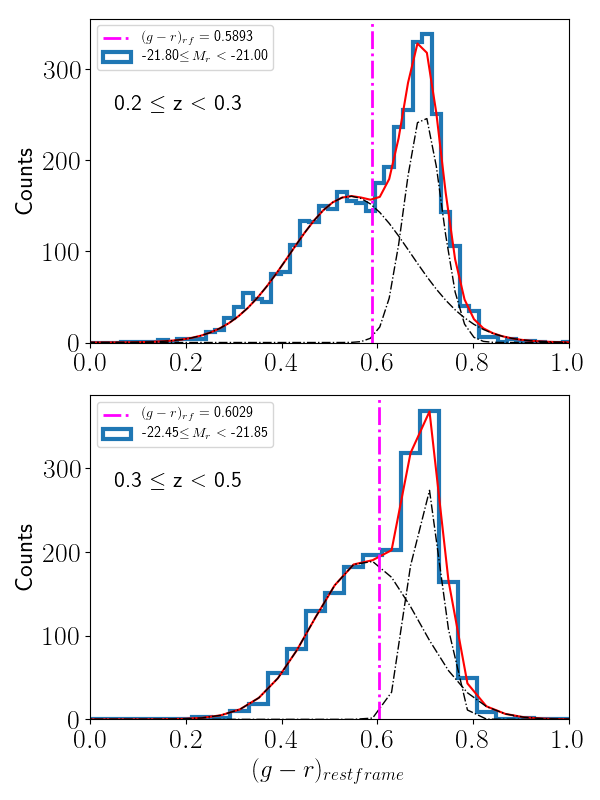}
\caption{Rest-frame (g-r) colour distributions performed in one representative absolute magnitude bin in the two highest redshift bins indicated in each panel. Curves and colours are shown as in Fig. \ref{CMD_fit_0102}.}
\label{CMD_fit_0205}
\end{center}
\end{figure}
To define the slope of the colour-magnitude cut, we focussed on the lowest redshift bin, which has a sufficiently wide magnitude range. We considered five 0.6 absolute magnitude bins and plot the $\rm (g-r)_{rest-frame}$ histogram of each subpopulation (Fig. \ref{CMD_fit_0102}). We then fit the histogram with a double-Gaussian curve and determined the minimum of the  distribution between the two peaks.
We computed the line interpolating the $\rm (g-r)_{rest-frame}$ colours just found in the five magnitude bins and used it to divide the galaxy population as shown in Fig. \ref{CMD_w1} (magenta dashed line).
At higher redshift, the magnitude range is too small to apply the same procedure. As no significant evolution is expected in the slope of the relation, but only in the zero point, we fixed the slope to that of the lowest z bin and computed the appropriate zero points with the same method outlined above (Fig. \ref{CMD_fit_0205}): we considered one magnitude bin at each redshift, we drew the $\rm (g-r)_{rest-frame}$ colour histogram and fit the distribution with a double-Gaussian curve, finding the local minimum between the two peaks.

To conclude, at 0.1$\leq$z$<$0.2 galaxies were assigned to the blue sequence if their colour obeys $(g-r)_{rest-frame}<-0.019 M_r+0.192$, at 0.2$\leq$z$<$0.3 the zero point is 0.177 and 0.176 at 0.3$\leq$z$\leq$0.5.

As a comparison between the two criteria just described we note that, considering all the redshift bins together, blue galaxies have a median sSFR$ \sim 10^{-9.7}$ $\rm yr^{-1}$ (and 90\% of galaxies have sSFR$\gtrsim10^{-10.45}$ $\rm yr^{-1}$).
Conversely, star-forming galaxies have a median $(g-r)_{rest-frame} \sim 0.58$ (and 90\% of galaxies have $(g-r)_{rest-frame} < 0.725$).

It is important to bear in mind that the two tracers used to characterise the galaxy populations have a different physical meaning and refer to different timescales.
While the SFR is an instantaneous measure of the rate at which a galaxy is forming stars at the epoch it is observed, colours are the result of longer processes tracing the predominant stellar population of a galaxy, whose colour is sensitive to its past history and to its current star formation activity. Moreover, colour is also influenced by other characteristics, such as the metallicity and the presence of dust.
In addition, the methodologies adopted to compute SFR and colours are different. The ongoing SFR is a product of the full spectral fitting analysis performed  on the spectra, while rest-frame colours are derived by means of SED fitting on the photometry.
Therefore, it is important to investigate the two quantities separately and study the incidence of each population over the total, as we do in the next sections.

\begin{table*}
\centering
\caption{Number of galaxies in the different environments (clusters in superclusters (S), clusters not in superclusters (NS), and field) and above the magnitude and mass completeness limits, in three redshift bins. Galaxies in clusters are further subdivided into virial and outer members.  The quantities in parentheses refer to the number of galaxies weighted for spectroscopic completeness. 
\label{tab:sampleW1}}
\setlength{\tabcolsep}{3pt}
\begin{tabular}{cl|cc|cc|cc|cc}
\multicolumn{2}{c|}{{\bf Environment}} &
\multicolumn{2}{c|}{$0.1 \leq z<0.2$} &
\multicolumn{2}{c|}{$0.2 \leq z<0.3$} &
\multicolumn{2}{c|}{$0.3 \leq z\leq0.5$}& 
\multicolumn{2}{c}{All}\\
 &&  $r \leq 20$ & $\rm M_\ast>10^{9.5}$ & $r \leq 20$ & $\rm M_\ast>10^{10.3}$  &$r \leq 20$ & $\rm M_\ast>10^{10.8}$ &$r \leq 20$ & $\rm M_\ast>M_{lim}$\\
\hline
\multicolumn{2}{l|}{Clusters (S)} &\multicolumn{2}{c|}{16} &\multicolumn{2}{c|}{20}&\multicolumn{2}{c|}{32}  & \multicolumn{2}{c}{68}\\
& N$_{gal}$ virial & 359 (571) & 348 (554) & 106 (146) & 99 (137) & 67 (152) & 57 (132) & 532 (869) & 504 (823)\\
& N$_{gal}$ outer & 454 (735) & 412 (670) & 136 (186) & 104 (142) & 100 (201) & 80 (165) & 690 (1122) & 596 (977) \\
\multicolumn{2}{l|}{Clusters (NS)} &\multicolumn{2}{c|}{8} &\multicolumn{2}{c|}{17}&\multicolumn{2}{c|}{18}  & \multicolumn{2}{c}{43}\\
&  N$_{gal}$ virial & 71 (142) & 65 (133) & 144 (191) & 122 (161) & 51 (67) & 46 (60) & 266 (400) & 233 (354) \\
& N$_{gal}$ outer & 103 (185) & 102 (183) & 193 (270) & 127 (182) & 57 (77) & 44 (56) & 353 (532) & 273 (421) \\
\multicolumn{2}{l|}{Field} &\multicolumn{2}{c|}{} &\multicolumn{2}{c|}{}&\multicolumn{2}{c|}{}  & \multicolumn{2}{c}{}\\
&  N$_{gal}$ & 5145 (9793) & 4511 (8577) & 6911 (10808) & 4615 (7181) & 4502 (7405) & 3125 (5180) & 16558 (28006) & 12251 (20938) \\
\end{tabular}
\end{table*}

\section{Results I: Galaxy population properties as a function of the {\it global} environment}
\label{sec:results_global}

In this section, we study the fractions and star-forming properties of galaxies in different {\it global} environments. 
We consider galaxies in the following environments. 

\begin{itemize}
\item Cluster virial members are galaxies whose spectroscopic redshift lies within $\rm 3 \sigma$ from the mean redshift of their host cluster, where $\rm \sigma$ is the velocity dispersion of their cluster and whose projected distance from the cluster centre is $< 1 \, r_{200}$.
\item Cluster outer members are galaxies whose spectroscopic redshift lies within $\rm 3 \sigma$ from the mean redshift of their host cluster, and whose projected distance from the cluster centre is between 1 and 3 $r_{200}$.
\item Galaxies in the field are all galaxies that do not belong to any cluster.
\end{itemize}

We note that all galaxies belonging to a structure are always included in the same redshift bin. For example, if a cluster is located at the edge of a redshift bin and its members spill over another bin, these are all included in the redshift bin of their host cluster, regardless of their actual redshift. 

We also treat separately virial and outer members that belong or do not belong to a supercluster.

Table \ref{tab:sampleW1} reports the number of galaxies in the different environments and redshift bins.
For all of these subsamples, numbers are given for the magnitude limited and mass limited samples.
At $0.1\leq z<0.2$ our sample includes three superclusters, at $0.2\leq z< 0.3$ three superclusters, and at $0.3\leq z\leq0.5$ six superclusters.

\begin{figure*}
\begin{center}
\includegraphics[scale=0.5]{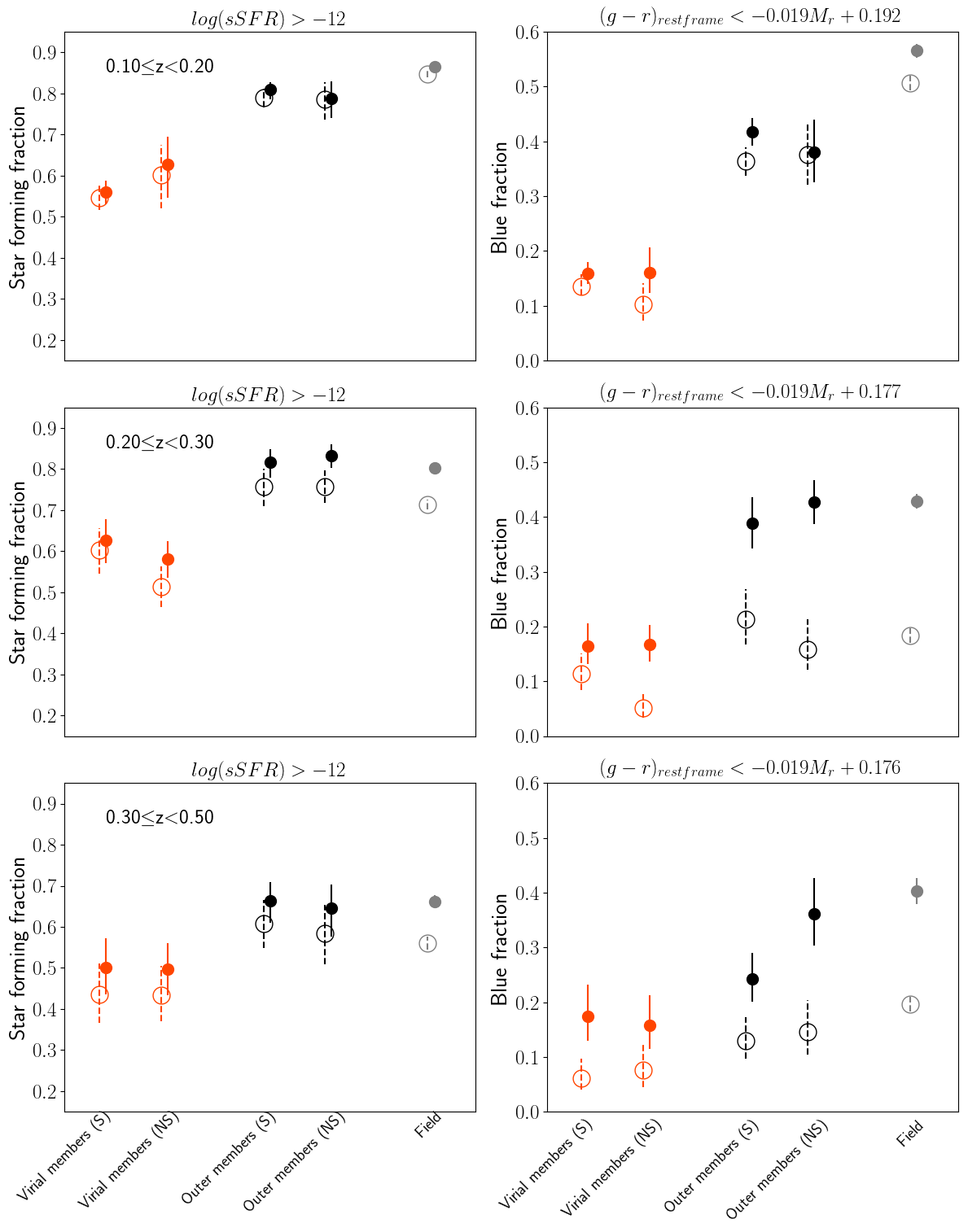}
\caption{Fraction of star-forming (left) and blue (right) galaxies in different environments and different redshifts, as indicated in the panels. Cluster members are divided into four subsamples: virial and outer members that belong or do not belong to a supercluster.
Values obtained using the magnitude limited sample are represented with filled symbols and solid errors, those obtained using the mass limited sample are represented by empty symbols and dashed error bars. A horizontal shift is applied for the sake of clarity. Errors are derived by means of a bootstrap method.}
\label{F_SFing_B_paper}
\end{center}
\end{figure*}

\subsection{Fraction of blue and star-forming galaxies}
\label{sec:SFing_blue_frac}

Figure \ref{F_SFing_B_paper} shows the fraction of blue and star-forming galaxies, separately, in the different global environments and in the three redshift bins, both for the magnitude limited and mass limited samples.
Error bars are computed using a bootstrap method. For galaxies in the field, we include in the error budget both the bootstrap error and the uncertainty due to the cosmic variance. Following \cite{Marchesini2009}, we sliced our field into nine right ascension subregions and we computed the fraction of star-forming and blue galaxies of each region separately; the contribution to the error budget from cosmic variance is then the standard deviation of the newly computed fractions divided by the number of subregions considered.

Overall, at all redshifts, both considering  the star formation and colours as tracers, fractions are similar within and outside the superclusters, suggesting that neither additional quenching processes nor triggering of the star formation are associated with the presence of superclusters.

At 0.1$\leq$z$<$0.2 (top left), both in the magnitude and in the mass limited samples, the star-forming fraction strongly depends on environment. 
Virial members have the lowest fraction of star-forming galaxies (55-60\%). This fraction increases when considering outer members, where $\sim 80\%$ of galaxies are star forming. Finally,  the percentage of star-forming galaxies in the field is the highest (86$\pm$1\%). The same trends are recovered when considering galaxy colours, even though fractions are systematically lower: $\sim 16\%$ of virial members are blue, as are $\sim 40\%$ of outer members and 57\% of field galaxies. Similarly to the star-forming fractions, results in the magnitude and mass limited samples are similar, except for the field value, where they differ by $\sim 10\%$; the mass limited sample shows a lower fraction than the magnitude-limited sample.

At 0.2$\leq$z$<$0.3 (middle panels of Fig. \ref{F_SFing_B_paper}), in both samples, virial members still show a significantly lower fraction of star-forming galaxies than the other environments ($\sim 55-60\%$), while outer members and field galaxies present very similar fractions ($\sim 85\%/75\%$ in the magnitude/mass limited samples). 
Considering colour fractions, the same trends are detected in the magnitude limited sample, where blue galaxies are $\sim 17\%$ in virial members, $\sim$42\% in outer members and in the field. In the mass limited samples, 
the difference between outer members and the field is  much smaller: the fraction of blue galaxies in these environments is always $<20\%$.

We recall that this redshift bin contains the XLSSsC N01 supercluster, separately discussed in XXL Paper XXX, and that contributes to the (S) cluster population with 11 out of 20 clusters, corresponding to $\sim 65\% $ of the cluster population.
In that supercluster an enhancement of the star formation activity of outer members with respect to the virial population and the field was observed. Nonetheless, general trends are maintained within the errors.

At 0.3$\leq$z$\leq$0.5 (bottom panels of Fig. \ref{F_SFing_B_paper}), both in the mass and magnitude limited samples, virial members 
have the lowest star-forming fraction (45-50\%), but differences with the other environments are reduced: in outer members and in the field the star-forming fractions are $\sim 65\%$ in the magnitude limited sample and $\sim 55-60\%$ in the mass limited sample. Considering colours, in the magnitude limited sample we still detect the usual differences between virial members and galaxies in  other environments, while in the mass limited sample all fractions are lower than $15\%$ and no variation with environment is detected.

As our cluster sample spans a wide range of X-ray luminosity (see Fig. \ref{cluster_fig}), we repeat the analysis separating the clusters in bins of  X-ray luminosity, but find no significant additional trends (plot not shown).

To summarise,  at all redshifts, field galaxies have the highest incidence of star-forming/blue galaxies, while virial members exhibit a noticeable suppression of both star-forming and blue fractions with respect to the other environments. Outer members exhibit a significant suppression of the star-forming/blue fractions with respect to the field only at 0.1$\leq z<$0.2, while at higher redshift they present similar fractions. 
No significant differences are detected between galaxies within and outside superclusters. However, fractional differences within and outside of superclusters do not follow a common trend at all redshifts, likely reflecting the variation of properties of individual supercluster structures at different redshifts.
The choice of a mass or magnitude limited sample only marginally affects the star-forming fractions, while it strongly alters those based on colours at $z>0.2$.

Overall, star-forming and blue fractions are never consistent within the errors: this is a probe that the two quantities, even though strictly related, are actually reflecting different aspects of the evolution of the galaxies. We note that in our sample no reasonable and physically motivated cut could be adopted to reconcile the fractions of star-forming and blue galaxies.

\begin{figure*}
\begin{center}
\includegraphics[scale=0.4]{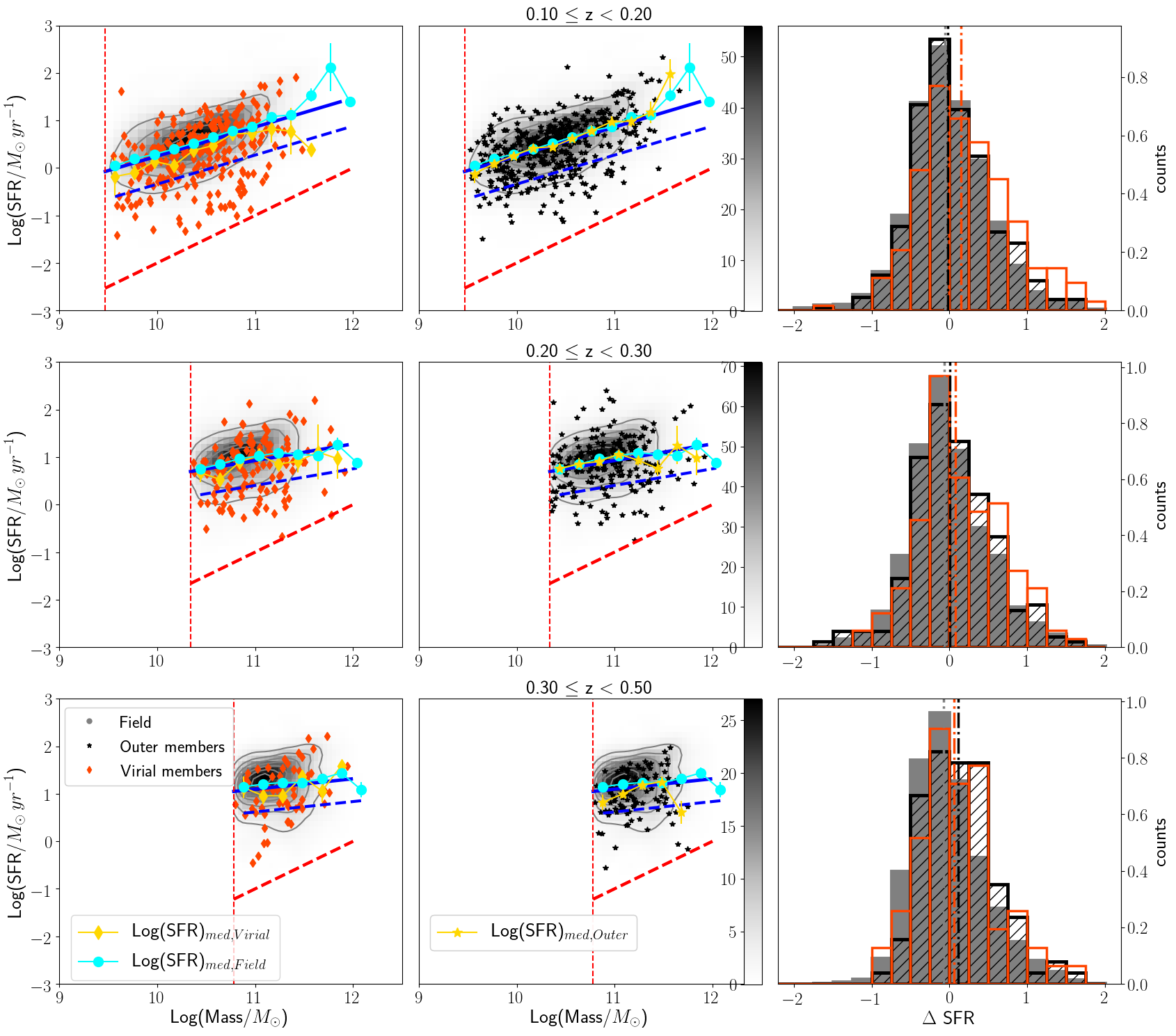}
\caption{\emph{Left and middle panels}. SFR-M$_\star$ relation for galaxies in the field and cluster virial and outer members (grey 2D histogram and density contours, orange diamonds, and black stars, respectively) in the mass limited sample. Panels in different lines refer to different redshift bins. The field population is represented with a 2D histogram whose values are given in the colour bar included in the middle panel, and grey contours trace the density levels of the data points. The vertical red dashed line shows the stellar mass limit at each redshift, while the oblique red dashed line sets the limit to the star-forming population, i.e. $sSFR=10^{-12}yr^{-1}$. The blue line is the linear fit to the SFR-M$_\star$ relation including all the environments at each redshift, and the dashed blue lines correspond to 1$\sigma$ errors on the fitting line. The parameters of the fit and the values of $\sigma$ are given in Table \ref{Table:SFR-M_fit}. The gold diamonds/stars and cyan dots represent the median SFR values computed in mass bins of 0.2 dex width, for the virial/outer members and field population, respectively. Error bars on the medians are computed assuming a normal distribution of the data points as 1.253$\sigma/\sqrt{n}$, where $\sigma$ is the standard deviation of the distribution and $n$ is the number of objects in the considered stellar mass bin. \emph{Right panels}. Histograms of the differences between the expected SFR computed using the main sequence fitting line at the stellar mass of any given galaxy in our sample and its actual SFR ($\Delta$SFR). Positive values of $\Delta$SFR indicate reduced SFR compared to the SFR main sequence of star-forming galaxies. The median values of the distributions are also shown with vertical dashed lines and different environments are colour coded as written in the legend.}
\label{SFR_M_z_tr}
\end{center}
\end{figure*}

In principle, the difference in the star-forming and blue fractions could be due to the presence of AGNs; for example, low-ionisation nuclear emission-line regions (LINERS) identified as red star-forming galaxies. These AGNs would increase the number of galaxies pertaining to the star-forming population without enhancing the fraction of blue galaxies. To test this, we removed broad- and narrow- line AGNs from our galaxy sample, as described in detail in Appendix \ref{app:AGN}, and we computed again the star-forming/blue fractions. The fractions are substantially unchanged (plot not shown), indicating that our results are not driven by the possible presence of AGNs.

We stress that comparisons across the different redshift bins are not possible, as magnitude and mass values used to define the sample are different. Furthermore, we point out that the decrease of the blue/star-forming fraction with increasing redshift is simply an artefact due to the galaxy mass range probed at different redshifts.

\begin{table*}[!t]
\centering
\caption{Fraction of galaxies in transition in different environments in the three redshift bins. Numbers are weighted for spectroscopic incompleteness and are computed above the stellar mass completeness limit of each redshift bin; the values in parenthesis refer to the highest stellar mass limit to allow comparisons at different redshifts. Errors are computed by means of bootstrapping. The last two lines of the table correspond to the values computed in two bins of LD and are analysed in Sect. \ref{sec:results_LD}.\label{tab:F_Transition}}
\hspace{-1.cm}
\begin{tabular}{l|c|c|c}
\hline
& $0.1 \leq z < 0.2$ & $0.2 \leq z < 0.3$ & $0.3 \leq z \leq 0.5$ \\
& $\rm log(M/M_\odot) >$9.5 & $\rm log(M/M_\odot) >$10.3 & $\rm log(M/M_\odot) >$10.8 \\
\hline
Field & $0.12^{+0.01}_{-0.01}$ ($0.13^{+0.02}_{-0.01}$) & $0.15^{+0.01}_{-0.01}$ ($0.16^{+0.01}_{-0.01}$) & $0.16^{+0.01}_{-0.01}$  \\
Cluster virial members &  $0.24^{+0.03}_{-0.03}$ ($0.24^{+0.06}_{-0.05}$) & $0.27^{+0.04}_{-0.04}$ ($0.24^{+0.05}_{-0.04}$) & $0.20^{+0.07}_{-0.05}$ \\
Cluster outer members & $0.15^{+0.02}_{-0.02}$ ($0.13^{+0.05}_{-0.04}$) & $0.17^{+0.03}_{-0.02}$ ($0.18^{+0.03}_{-0.03}$) & $0.18^{+0.05}_{-0.04}$\\
\hline
Cluster virial members (S) & $0.27^{+0.03}_{-0.03}$ ($0.25^{+0.08}_{-0.07}$) & $0.21^{+0.06}_{-0.05}$ ($0.17^{+0.06}_{-0.05}$) & $0.15^{+0.10}_{-0.06}$ \\
Cluster virial members (NS) & $0.15^{+0.06}_{-0.05}$ ($0.21^{+0.13}_{-0.09}$) & $0.32^{+0.06}_{-0.05}$ ($0.32^{+0.07}_{-0.06}$) & $0.30^{+0.10}_{-0.08}$\\
Cluster outer members (S) & $0.15^{+0.02}_{-0.02}$ ($0.13^{+0.04}_{-0.03}$) & $0.10^{+0.04}_{-0.03}$ ($0.11^{+0.04}_{-0.03}$) & $0.23^{+0.07}_{-0.05}$\\
Cluster outer members (NS) & $0.15^{+0.07}_{-0.05}$ ($0.14^{+0.14}_{-0.18}$) & $0.22^{+0.04}_{-0.04}$ ($0.23^{+0.05}_{-0.05}$) & $0.07^{+0.06}_{-0.03}$\\
\hline
High-LD (85th) & $0.13^{+0.02}_{-0.01}$ ($0.15^{+0.04}_{-0.03}$) & $0.15^{+0.02}_{-0.01}$ ($0.15^{+0.02}_{-0.02}$) & $0.13^{+0.02}_{-0.02}$\\
Low-LD (15th) & $0.14^{+0.02}_{-0.01}$ ($0.12^{+0.04}_{-0.03}$) & $0.14^{+0.01}_{-0.01}$ ($0.16^{+0.02}_{-0.02}$) & $0.17^{+0.03}_{-0.02}$ \\
\hline
\end{tabular}
\end{table*}

\subsection{SFR-mass relation}
\label{sec:sfr_m_glob}

We focus in this section only on the star-forming population and investigate the correlation between the SFR and galaxy stellar mass (SFR-M$_\star$). 
For this analysis we only rely on the mass limited sample. Indeed, in contrast with the magnitude limited sample, applying a mass limit ensures completeness, i.e. to include all galaxies more massive than the limit regardless of their colour or morphological type.
This ensures that we do not bias the results because of the absence of galaxies which are under-sampled or missed by selection effects, as might happen when considering a magnitude limited sample.
As in the previous section we did not detect any significant difference between galaxies within and outside superclusters, in what follows we do not distinguish between the two subgroups.

Figure \ref{SFR_M_z_tr} compares the distribution of galaxies in different environments and in different redshift bins in the SFR-M$_\star$ plane (left and middle panels). Roughly, at all redshifts, galaxies located in the different environments share a common region on the plane, excluding strong environmental effects at play. Comparing the galaxies at different redshifts, we find a decline in SFR with time at fixed stellar mass, in agreement with many previous literature results \citep[e.g.][]{Noeske2007,Vulcani2010}.

To probe the apparent lack of environmental effects on a statistical ground, we proceed by first performing a  linear regression fit to the relation by considering all the different environments together and then compare the median values of SFR in different mass bins for the various
environments to this fit. The values of the best-fit slope, intercept and 1$\sigma$ are given in Table \ref{Table:SFR-M_fit}. Error bars on the medians are computed in each stellar mass bin as 1.253$\sigma/\sqrt{n}$, where $\sigma$ is the standard deviation of the SFR distribution in the bin and $n$ is the number of objects considered in the bin.

\begin{table}
\centering
\caption{Best-fit parameters of the linear fit to the SFR-M$_\star$ relations shown in Fig. \ref{SFR_M_z_tr}, in three redshift bins. The fit is performed on the sample including all the environments together, and the fitting line has the following general equation: Log(SFR)= $a$Log(M$_\star$)+$b$. \label{Table:SFR-M_fit}}
\begin{tabular}{c|c|c|c}
\hline
& $a$ & $b$ & $\sigma$ \\
\hline
0.1$\leq$z$<$0.2 & 0.61 & -5.86 &  0.59\\
0.2$\leq$z$<$0.3 & 0.35 & -2.92 & 0.52\\
0.3$\leq$z$<$0.5 & 0.22 & -1.31 & 0.48\\
\hline
\end{tabular}
\end{table}


The fit to the SFR-M$_\star$ relation is dominated by field galaxies, whose median trends closely follow the fitting line at all redshifts. In contrast, cluster virial members show hints of lower median SFR with respect to the latter in all the redshift bins; some statistical oscillations are due to the lower number of galaxies at 0.3$\leq$z$\leq$0.5. Furthermore, in this case the limited mass range could also affect the reliability of the fit. The median SFR of outer members closely follows the field trend at z$\leq$0.2 and is compatible within the error bars with both the field and virial members at higher redshift. We do not plot these values for the sake of clarity.

The right-hand panels of Fig. \ref{SFR_M_z_tr} report the distribution of the differences between the SFR of each galaxy and the value derived from the global fit given the galaxy mass ($\Delta$SFR), for any given environment. Positive values of $\Delta$SFR correspond to reduced SFR with respect to the expected value.
At all redshifts, it is immediately clear that the shape of distribution of $\Delta$SFR of virial members differs from that of the field population, whereby the former presents a tail of reduced SFR values with respect to the latter. 
A Kolmogorov-Smirnov (KS) test is able to detect differences between virial members and field galaxies at all redshifts (P(KS) $\leq 0.05$); outer members instead have statistically different distributions with respect to the field only at 0.3$\leq$z$\leq$0.5 (P(KS)$<$0.02), and with respect to virial members only at 0.1$\leq$z$<$0.2 (P(KS)$<$10$^{-3}$). Nonetheless, at all redshifts, median values are compatible within the errors among the different samples, indicating that the tail, although present in virial and outer members, is not able to affect the whole SFR distribution significantly.

\subsection{Galaxies in transition}
\label{sec:F_TR_global}

The presence of a non-negligible number of galaxies with reduced SFR among the cluster population motivates a more detailed investigation on the presence of the so-called galaxies in \emph{transition}, i.e. star-forming galaxies which are slowly decreasing their SFR and are detected as an intermediate population migrating from the star-forming main sequence down to the quenched population. 
To identify the galaxies in transition we follow \cite{Paccagnella2016}, and select galaxies with $\rm (sSFR)>10^{-12} yr^{-1}$ and SFR below 1$\sigma$ from the SFR-M$_\star$ fitting line.
The transition fraction is computed as the ratio of this population to the number of star-forming galaxies in each environment.
We note that, by definition, the percentage of galaxies below a 1$\sigma$ cut of the SFR-M$_\star$ relation should be $\sim$15-17\%, therefore the identification of a population of galaxies in transition is measured as an excess of galaxies compared to this statistical value.

\begin{figure*}
\centering
\includegraphics[scale=0.47]{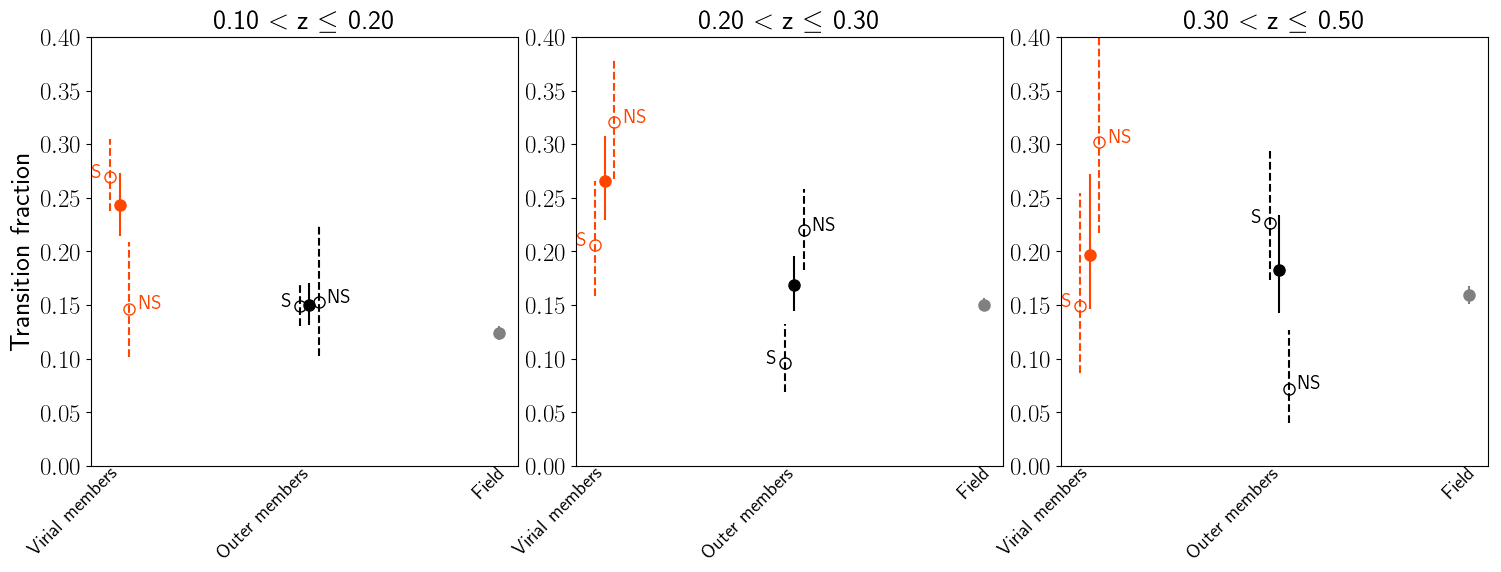}
\caption{Fraction of galaxies in transition in the mass limited sample in the three redshift bins. Filled dots represent galaxies in the different environments, as written in the x-axis. The (S) and (NS) contribution to the virial and outer member populations are also represented with empty symbols and dashed error bars. Error bars are computed via bootstrapping.\label{Fig:F_TR}}
\end{figure*}

The fractions of galaxies in transition as a function of environment for
different redshift bins are presented in Fig. \ref{Fig:F_TR} and given in Table \ref{tab:F_Transition}.
We compute these fractions also dividing virial/outer cluster members residing or not in superclusters.

The incidence of the population of galaxies in transition depends on environment. As shown in Fig. \ref{Fig:F_TR}, the fraction of transition galaxies
in the field and outer members is (within the errors) almost  half of that observed in cluster virial members at z$\leq$0.3. At higher redshift instead, the fractions are similar within the error bars in all environments, likely owing to the high stellar mass limit considered.

Considering separately clusters within and outside superclusters, no clear trends are observed in the transition fractions, suggesting again that differences among superclusters are most likely statistical. In this context, we note that at 0.2$\leq$z$<$0.3 the fraction of galaxies in transition in the virial and outer regions of (S) clusters is in agreement with the trends found for the XLSSsC N01 supercluster (XXL Paper XXX). The transition fractions are $\sim$ 10\% lower in both (S) virial and outer members compared to their (NS) counterparts, as in the XLSSsC N01 supercluster where the percentage of galaxies with reduced SFR was $<$20\% in all the environments.

We also tested whether the X-ray luminosity played a role in the determination of the number of galaxies in transition in clusters, and we did not find any clear correlation in the luminosity range probed by our cluster sample.

\begin{figure*}
\begin{center}
\includegraphics[scale=0.5]{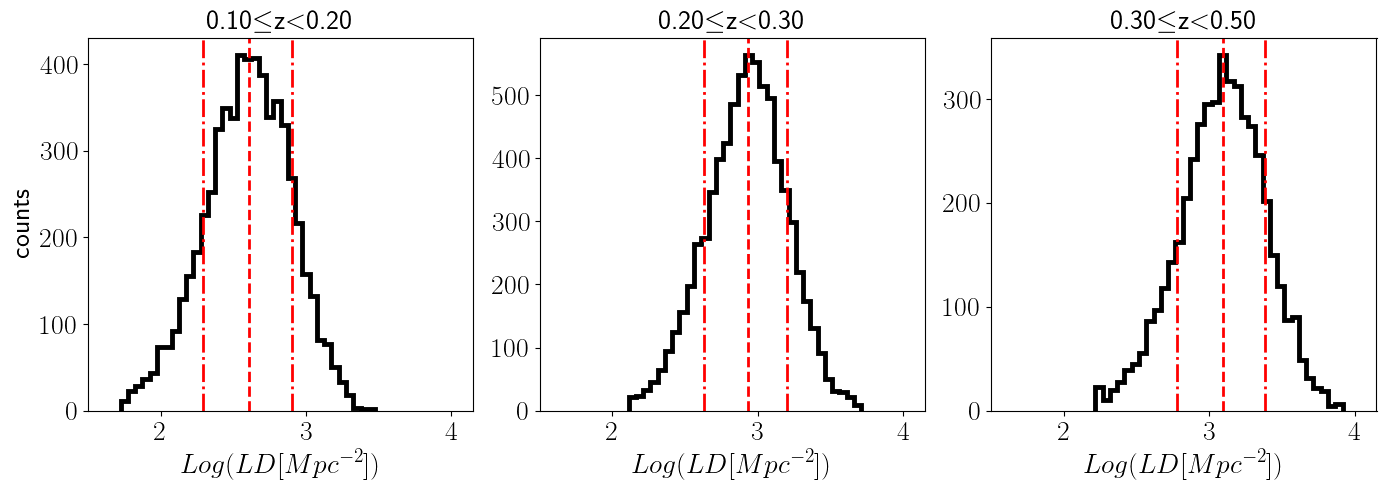}
\caption{Distributions of the logarithm of the LD in the three redshift bins, as indicated in the labels. Histograms are drawn after a sigma-clipping has been performed on the parent distributions. The red dashed vertical lines represent the 15th, 50th and 85th percentiles, respectively.
\label{histo_ld_w1}}
\end{center}
\end{figure*}

As a general understanding, environmental effects seem to dominate within the cluster virial radii: the substantial difference in the number of galaxies with reduced SFR among cluster virial members compared to the field population is responsible for detection of tails in the $\Delta$SFR distributions, shown in the right panels of Fig. \ref{SFR_M_z_tr}. 

\section{Results II: Galaxy population properties as a function of the {\it local} environment}
\label{sec:results_LD}

The availability of a large spectrophotometric sample of galaxies enables the parametrisation of environment also in terms of projected LD of galaxies.
In this section we consider together the galaxies in all the aforementioned environments and divide these sources into the usual three redshift bins. For each galaxy, we compute the projected LD as the number of galaxies enclosed into a fixed radial aperture of 1 Mpc at the redshift of the galaxy and within a given redshift range around the centre galaxy.
We describe the computation of LD in detail in Appendix \ref{LD_computation}.
Figure \ref{histo_ld_w1} shows the LD distribution in the three redshift bins in logarithmic units, along with the 15th, 50th, and 85th percentiles, which will be used to define the LD bins used in Sect.\ref{sec:LD_SFR_m}. It is evident that going from low- to high-z the peak (i.e. the median) of the LD is shifted towards higher densities, as previously found in other samples \citep{Poggianti2010}.

\subsection{Fraction of blue and star-forming galaxies}
\label{sec:LD_SFing_blue_frac}

\begin{figure*}
\begin{center}
\includegraphics[scale=0.5]{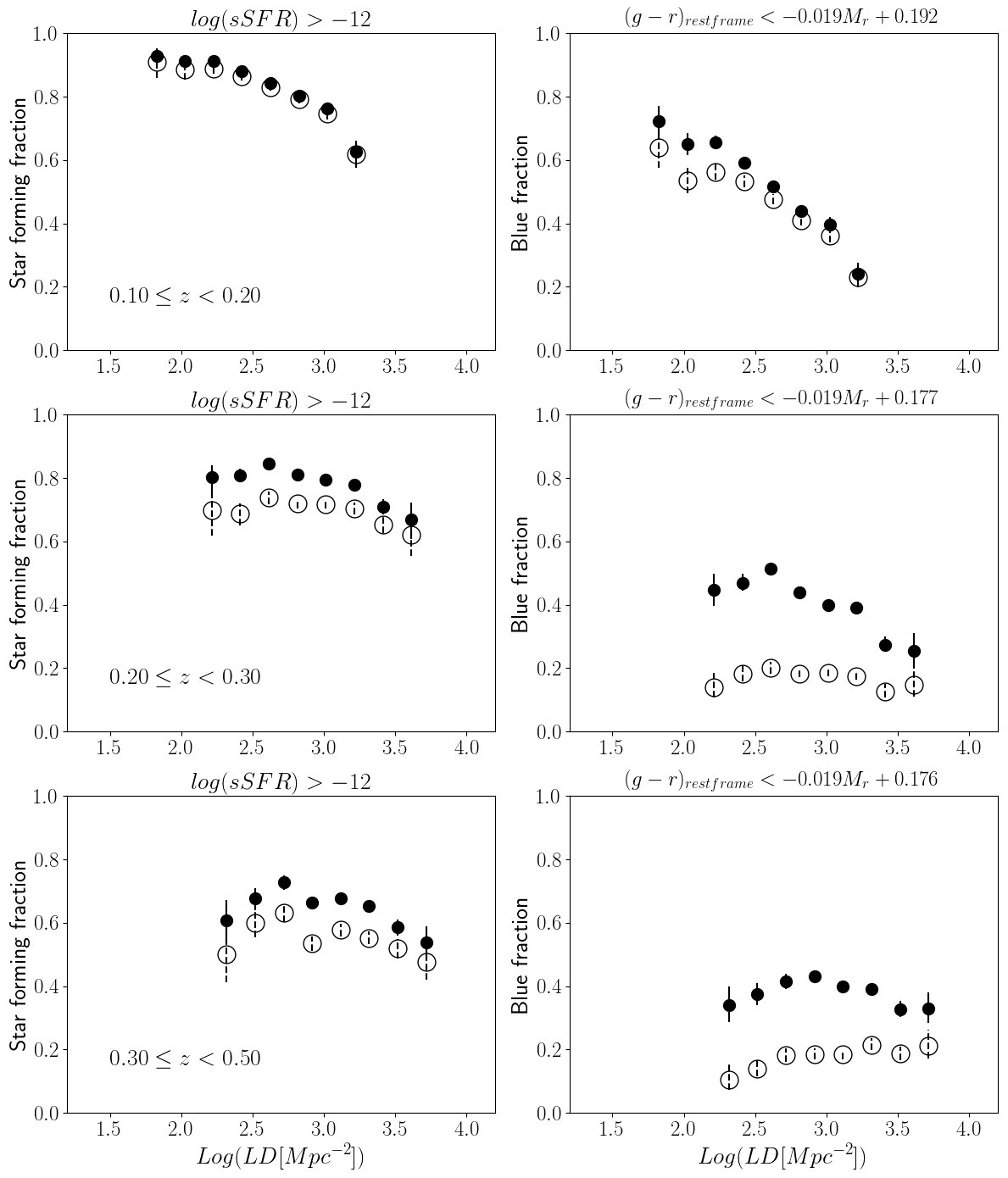}
\caption{Fraction of star-forming galaxies in different bins of LD, computed with the sSFR (left panels) and rest-frame colour (right panels). Three redshift bins from z=0.1 up to z=0.5 are represented, and the redshift increases from top to bottom panels as indicated in each panel. A sigma-clipping has been performed on the parent LD distributions to remove outliers and bins with a non-statistically representative number of objects. Panels and symbols are shown as in Fig. \ref{F_SFing_B_paper}.}
\label{SF_LD_0105}
\end{center}
\end{figure*}

Figure \ref{SF_LD_0105} shows the fraction of blue (right) and star-forming (left) galaxies as a function of the projected LD, in the three redshift bins, separately, for both the magnitude and mass limited samples. Error are derived by means of bootstrapping. 
At $0.1 \leq z<0.2$ (top panels), both in the magnitude and in the mass limited samples, the fraction of both star-forming and blue galaxies decreases monotonically with increasing LD. The star-forming fraction is close to 90\% at low densities and then decreases of a factor $\gtrsim$1.5 in a LD range of 2.0 dex; the blue fraction is $\sim 80\%$ at low densities and decreases of almost four times; the values drop to $\sim$0.2 at the highest densities.

\begin{figure*}
\begin{center}
\includegraphics[scale=0.35]{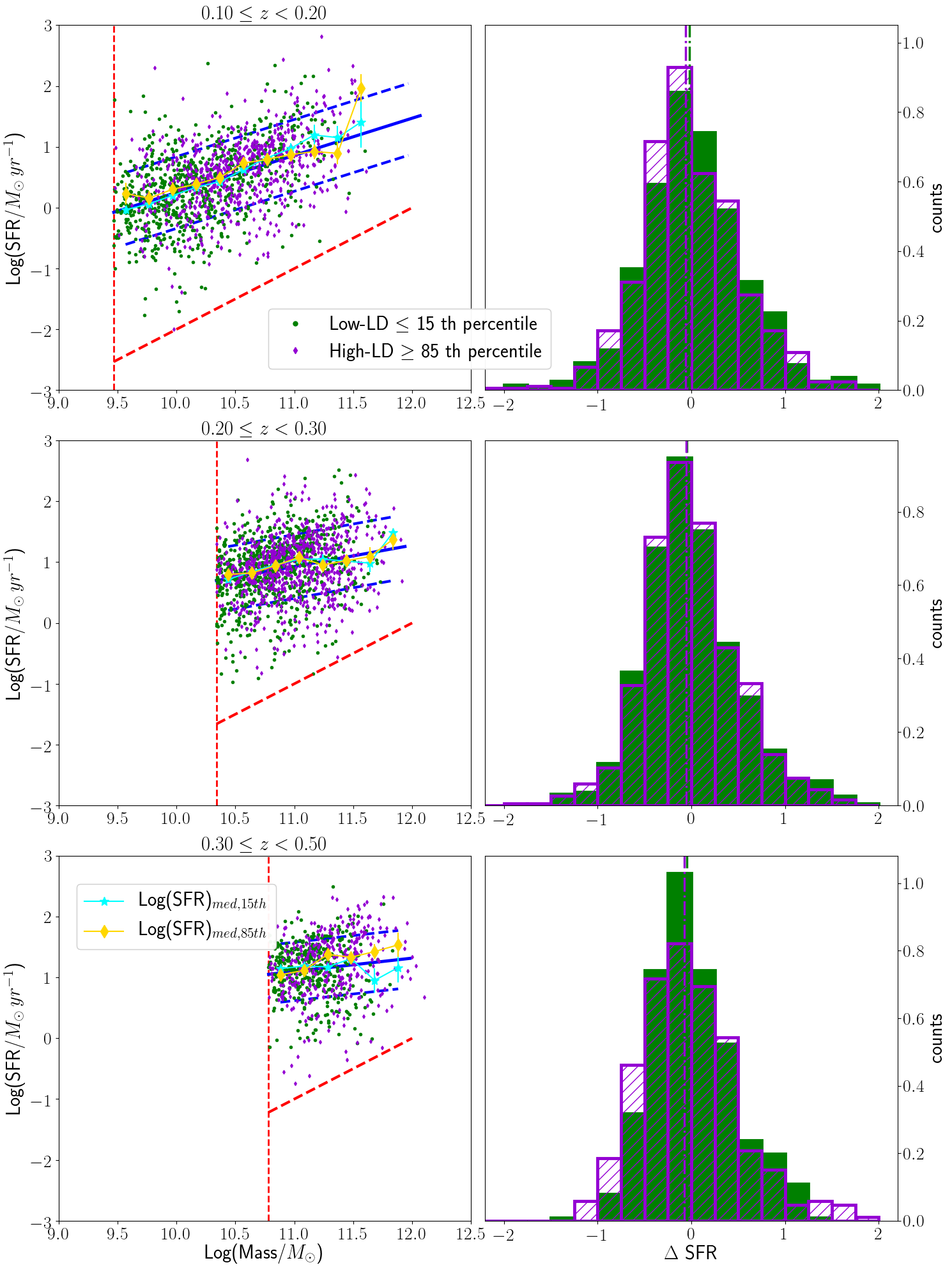}
\caption{\emph{Left panels}. SFR-M$_\star$ relation for galaxies in two regimes of LD, corresponding to the wings of the LD histograms shown in figure \ref{histo_ld_w1}. Panels and lines are shown as Fig. \ref{SFR_M_z_tr}.
Cyan stars and the gold diamonds represent the median values of the SFR computed in 0.2 dex stellar mass bins, for the low- and high- LD regimes respectively. Error bars are computed as in Fig. \ref{SFR_M_z_tr}.
\emph{Right panels}. Histograms of the differences between the expected SFR computed using the main sequence fitting line at the stellar mass of any given galaxy in our sample and its actual SFR ($\Delta$SFR).
Median values of the distributions are shown with vertical dashed lines and colour coded as written in the legend.}
\label{sfr_m_ld_diff}
\end{center}
\end{figure*}

At $0.2\leq z<0.3$ (middle panels of Fig. \ref{SF_LD_0105}), the star-forming fractions are much less dependent on density, both in the mass and magnitude limited samples. Values range between 80 and 60\%, at low and high density, respectively. 
In contrast, in the magnitude limited sample, the blue fraction still shows a significant decrease with LD, ranging from 50\% at low densities to 20\% at the highest. In the mass limited sample the blue fraction is always $\lesssim 20\%$, regardless of density. 

In the highest redshift bin (bottom panels of Fig. \ref{SF_LD_0105}), both in the magnitude and mass limited samples the star-forming fractions seem first to increase with density, reach a plateau and then decrease at the highest values. Overall, values range between 50 and 70\% in the magnitude limited sample, 40\% to 60\% in the mass limited sample. 
Such increase with LD is also noticeable in the colour fractions: in the magnitude limited sample at low density the fraction is $\sim 25\%$, reaches 40\% at intermediate densities and falls down to 30\% at the highest density. In the mass limited sample, the fraction of blue galaxies is always $<20\%$, but shows a statistically meaningful increase from the lowest to the highest densities. 

To summarise, the star-forming/blue fraction of galaxies decreases at densities higher than the LD median at each redshift (see Fig. \ref{histo_ld_w1}). At densities lower than the median, we notice a steady decrease of the fractions at 0.1$\leq$z$<$0.2, opposed to an initial increase at z$\geq$0.2. Furthermore, the overall decrease of the star-forming/blue fractions going from the low to high densities is much more pronounced at lower than at higher redshifts.
As it was previously found in Sect. \ref{sec:SFing_blue_frac}, considering either the magnitude limited sample or the mass limited sample lead to substantial differences only in the fraction of blue galaxies at z$>$0.2.
Finally, differences in the absolute values of star-forming and blue fractions are again noticeable and are further investigated and discussed in Sect. \ref{sec:Discussion_SFing_vs_blue}.

\subsection{SFR-mass relation and galaxies in transition}
\label{sec:LD_SFR_m}
We now study the SFR-M$_\star$ relation of galaxies in two extreme bins of LD representative of the lowest and highest LD environments.
With reference to the histrograms represented in Fig. \ref{histo_ld_w1}, we selected two percentiles that allowed us to seize the wings of the distribution (having previously removed outliers), considering its narrow shape. The selected percentiles are 15th and 85th.

In Figure \ref{sfr_m_ld_diff} we report the SFR-M$_\star$ relation of galaxies in the low- and high- LD regimes. We proceed as before and compute the median SFR in stellar mass bins of 0.2 dex width in the low- and high- LD regimes.
The median values of the SFR computed in bins of stellar mass show little variation with LD (yellow diamonds versus cyan stars), whose values that are always consistent within the error bars. Differences arising at the highest stellar mass values at z$\geq$0.3 may be mostly driven by the low sample statistics, and therefore should be taken with caution.
The right-hand panels of Fig. \ref{sfr_m_ld_diff} show the $\Delta$SFR with respect to linear fit to the SFR-M$_\star$ relation used in Sect. \ref{sec:sfr_m_glob}, computed as previously done for the global environment. 
The median $\Delta$SFR values are very similar in the high- and low-LD regimes at all redshifts, and the statistical similarity between the two samples is further confirmed by the outcome of the KS test: P(KS)$>>$0.05 at all redshifts.

Finally, we also compute the fraction of transition galaxies in the two extreme LD bins  (see Tab. \ref{tab:F_Transition}), finding no differences within the error, at all redshifts. 

\section{Discussion}
\label{sec:discussion}
In this paper we have adopted two definitions of environment. The first is based on the X-ray selection of virialised structures; the second is based on the local galaxy number density. We are now in the position of contrasting the results, and we aim to understand whether the different parametrisations lead to similar conclusions.

In the literature, the environmental dependence of the galaxy properties was previously investigated by many authors, adopting either a global or local parametrisation, but hardly ever directly contrasting the two in homogeneous samples. Nonetheless, as discussed by \cite{Vulcani2011,Vulcani2012,Vulcani2013} and \cite{Calvi2018}, the two definitions are not interchangeable and can give opposite results, highlighting that different processes dominate at the different scales probed by the different definitions.

As far as galaxy fractions are concerned, we find that regardless of the environmental definition adopted the fraction of blue/star-forming galaxies is systematically higher in the field/least dense regions than in the virial regions of clusters/highest densities. This effect is less significant in the highest redshift bin analysed. Our results are overall in line with what was previously found in the literature, both considering the global  \citep[e.g.][]{Iovino2010, Muzzin2012} and local \citep[e.g.][]{Balogh2004a, Cucciati2017} environments.
Similarly, the overall SFR-M$_\star$ relation also seems not to depend on the parametrisation adopted, which agrees with numerous literature results that claim the invariance of SFR-M$_\star$ relation on environment \citep[e.g][]{Peng2010}. 

Nonetheless, the two definitions of environment lead to different results when we analysed the fraction of galaxies in transition.
In fact, using the local environment the fraction of galaxies below the main sequence is similar at low and high density, whereas in clusters (and especially in their virial regions) a population with reduced SFR with respect to the field is observed.
This population is most likely in a transition phase of star formation and, although clearly detected, it is not able to affect the whole SFR-M$_\star$ relation because it constitutes a small fraction of all galaxies, as shown in Tab. \ref{tab:F_Transition}.

\subsection{Galaxies in transition in the different environments and their evolution with redshift}
\label{sec:discussion_F_tr}

The presence of a population of galaxies in transition from being star forming to passive was already detected in galaxy clusters by several works at low and intermediate redshifts \citep{Patel2009,Vulcani2010,Paccagnella2016}, and has been interpreted as an evidence for a slow quenching process preventing a sudden relocation of galaxies from the star forming to the red sequence.

\begin{figure}
\begin{center}
\includegraphics[scale=0.5]{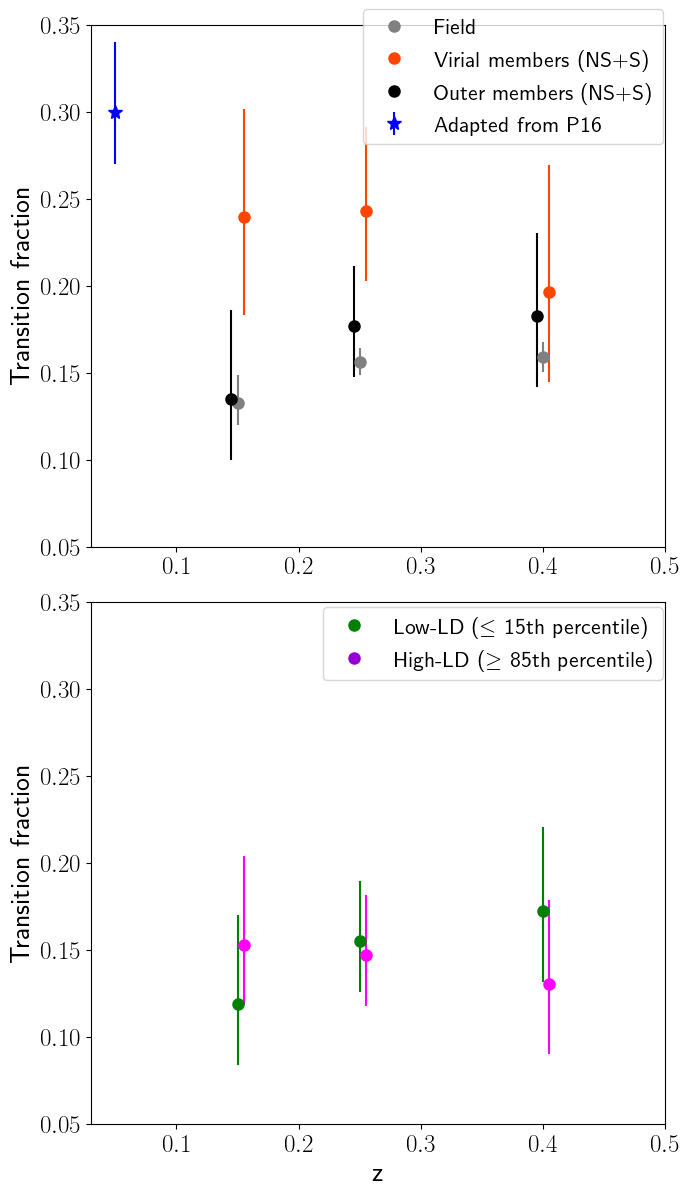}
\caption{Fraction of galaxies in transition at 0.1$\leq z \leq 0.5$ considering the global (top) and local (bottom) parametrisation 
Fractions are computed for $\log M/M_\odot \geq 10.8$, the stellar mass completeness limit at 0.3$\leq$z$\leq$0.5. Error bars on the fractions are computed via bootstrapping. In the top panel, the blue star represents the fraction of transition galaxies in the local universe, adapted from \citet{Paccagnella2016}.}
\label{F_trans_Mlim}
\end{center}
\end{figure}

In the previous sections, it was not possible to investigate the evolution of the incidence of transition galaxies, as a different mass complete limit was adopted at each redshift. 
Now we consider instead the same mass limit, to allow for fair comparisons. We adopt the most conservative value, that is the mass completeness limit in the highest redshift bin.
Fractions are given in parenthesis in Tab. \ref{tab:F_Transition}.

Figure \ref{F_trans_Mlim} shows the fraction of galaxies in transition in the redshift range 0.1$\leq$z$\leq$0.5 considering the global and local environments.

The upper panel shows that in the case of global environment the overall fraction of transition galaxies with $\log (M_\star/M_\odot)>10.8$  does not significantly vary with cosmic time, remaining around $\sim 15$\%, both in the field and among outer members. In contrast, virial members present higher transition fractions with a tentative increase as time goes by, although uncertainties prevent us from drawing solid conclusions.

The same Figure also compares our results to those obtained at low redshift (z$\lesssim$0.1) by \cite{Paccagnella2016}, when the subsample of their cluster galaxies within 1$r_{200}$ and with stellar masses $\rm M_\star \geq 10^{10.8} M_\odot$ is considered.
The resulting transition fraction weighted for incompleteness is 0.30$^{+0.04}_{-0.03}$, that is consistent with our results within the error bars and point towards the aforementioned increase in the transition fractions at more recent epochs.

In contrast, the lower panel of Fig. \ref{F_trans_Mlim} shows no dependence of the transition fraction with redshift for galaxies located at different local densities, further demonstrating that the local environment does not affect the incidence of such population.

Evidently, the two parametrisations are able to probe different physical conditions for galaxies, determining different timescales in the star formation process and quenching timescales.

\subsection{Star-forming versus colour fractions in the different environments}
\label{sec:Discussion_SFing_vs_blue}

\begin{figure} 
\centering
\includegraphics[scale=0.6]{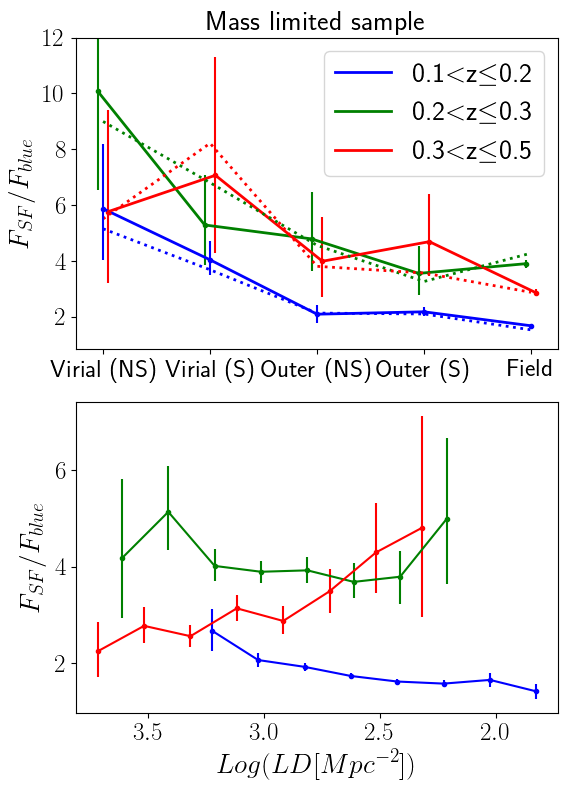}
\caption{Ratio of the fraction of star-forming (F$_{SFing}$) to blue (F$_{blue}$) galaxies in the mass limited sample in the three redshift bins and in different global (top) and local (bottom) environments.
Dashed lines in the top panel show trends when AGNs are removed form the  sample as explained in the Appendix \ref{app:AGN}.
In both panels, error bars are computed by propagating the asymmetric errors on the single fractions by means of the statistical error propagation. 
\label{Fig:F_SFing_Blue_ratio}}
\end{figure}

In the previous sections we have separately analysed the dependence of the star-forming and blue galaxy fractions on the global and local environments. 
In both analyses a difference between the star-forming and blue fractions emerged, wherein the former is systematically higher than the latter. 
We stress that this difference is not likely to be due to the definition we adopted for determining the two populations: as previously described in Sect. \ref{sec:pops}, the sSFR and colour threshold adopted for defining the star-forming and blue populations are physically motivated by the distribution of the galaxy samples in the sSFR-M$_\star$ plane and by the rest-frame colour distribution at different redshifts. We further explored whether a choice of different cuts either on the sSFR and on the (g-r) rest-frame colour led to more similar galaxy fractions and concluded that the resulting sSFR and/or colour threshold to apply to the population in order to reconcile the fractions were totally non-physical.

As already anticipated in Sect. \ref{sec:pops}, the two quantities present intrinsic differences related to the tracers they are based on: the SFR is derived from the measure of the flux of emission lines sensitive to the short-lived massive stars, while avoiding as much as possible contributions from evolved stellar populations. It is basically able to probe the presence of newly or recently formed stars on timescales of $\sim$10-100 Myr. On the contrary, galaxy integrated colours are more sensitive to the integrated star formation history and in particular to the stellar populations dominating the galaxy light, and are further influenced by the dust content and metallicity of the galaxy.
With this in mind, we can expect a good agreement between galaxy rest-frame colours and SFR indicators when the galaxy is actively forming stars at a steady rate on the main sequence or, conversely, when it is quiescent and has been passively evolving for some Gygayear. Differences between the two tracers may be expected for example when the galaxy suddenly interrupts its star formation activity as a consequence of the interactions with external physical mechanisms (e.g. environmentally related phenomena).

We are now in the position of directly comparing the fraction of star-forming and blue galaxies with the intent of obtaining some clues regarding the physical processes occurring in the different environments. 

Figure \ref{Fig:F_SFing_Blue_ratio} shows the ratio of the number of star forming to that of blue galaxies as a function of global (top panel) and local (bottom panel) environment, above the stellar mass completeness limit of each redshift bin.
In the upper panel of the figure, a strong dependence of the $F_{SFing}/F_{blue}$ ratio on the global environment emerges.
At 0.1$\leq$z$\leq$0.3, this ratio is highest in the virial regions of clusters, while it decreases in the other environments with little difference found between cluster outskirts and the field. Moving towards higher redshift, uncertainties prevent us from drawing solid conclusions, but still a hint of a higher $F_{SFing}/F_{blue}$ ratio within the virial radii of clusters than the other environments is visible.

In principle, this result might be contaminated by the presence of AGNs, and in particular LINERS, that could be misclassified as red star-forming galaxies. 
The dashed lines in Fig. \ref{Fig:F_SFing_Blue_ratio} show the $F_{SFing}/F_{blue}$ ratios after AGNs have been removed (see  Appendix \ref{app:AGN}) and that this population cannot be responsible for the observed trends. 

Our results suggest that in the innermost regions of clusters, besides the suppression of the star formation activity, further environmentally related physical processes come into play to produce a population of galaxies with a non-negligible SFR that however is not coupled with (blue) rest-frame colours.

This decoupling is most likely due to the different star formation histories that characterise galaxies in the different global environments. 
Indeed, \cite{Guglielmo2015} found that the star formation history of low-redshift star-forming galaxies has been decreasing since z$\sim$2, and in particular the rate at which stars were produced in galaxies in clusters at high-z is higher than in the field, regardless of their stellar mass. This implies that, on average, star-forming galaxies in clusters formed the bulk of their stellar mass at older epochs than their counterparts in the field. Thus these star forming galaxies host older stellar populations which have redder colours, although these galaxies still are forming stars at the epoch of observation. 

Alternatively, the presence of a population of red star-forming galaxies may be also associated with a dust obscured star formation phase.
\cite{Gallazzi2009} quantified that nearly 40\% of the star-forming galaxies in a supercluster at z$\sim$0.17 (Abell 901/902) had red optical colours at intermediate and high densities. These red systems have sSFR similar to or lower than blue star-forming galaxies, thus they are likely undergoing gentle mechanisms that perturb the distribution of gas inducing star formation (but not a starburst) and at the same time increase the gas/dust column density.

The incidence of the red star forming population is instead less dependent on the local environment: the lower panel of Fig. \ref{Fig:F_SFing_Blue_ratio} shows no strong trends of  the $F_{SFing}/F_{blue}$ ratio with LD at any redshift, also because of the large uncertainties, especially at higher redshifts.  

These trends prove, once again, that the two environmental parametrisations are probing galaxies in different physical conditions, and that they cannot be used interchangeably. Indeed, there is no constant direct correspondence between the cluster cores and the highest LD regions and, similarly, between the lowest LD regions and the field.

\section{Conclusions}
\label{sec:conclusions}

In this work, we have conducted a study on the stellar population and star formation properties of galaxies in the range $0.1\leq z\leq 0.5$, by making use of two definitions of environment. When considering the {\it global} environment, we divided galaxies into cluster virial and outer members and the field. We also distinguished between clusters that belong or do not belong to a supercluster. 
When considering the {\it local} environment, we characterised galaxy properties as a function of the projected LD.

The main observables we considered for investigating galaxy properties in different environments are the fraction of star-forming/blue galaxies, defined on the basis of the sSFR and colour, respectively, and the correlation between the SFR and stellar mass.
The main results can be summarised as follows. 

\subsubsection*{\it Fraction of star-forming and blue galaxies} Considering the global environment, both in the magnitude and in the mass limited samples, cluster virial members reveal a deficiency of star-forming/blue galaxies with respect to all other environments at all redshifts, while field galaxies are the most star-forming/blue population at all redshifts. Outer members exhibit a significant suppression of the star-forming/blue fractions with respect to the field only at 0.1$\leq z<$0.2, while at higher redshift they present similar fractions.
Overall, no significant differences are detected between galaxies within and outside superclusters. 

Considering the LD instead, the star-forming/blue fraction steadily decreases with increasing density only at 0.1$\leq$z$<$0.2. At higher redshift, the fractions show a qualitatively similar dependence on density for $\log(LD [Mpc^{-3}])\gtrsim$3, while at lower densities the trends slightly increase.  

Regardless of the parametrisation of the environment, star-forming and blue fractions are never consistent within the errors, probing that the two quantities reflect different aspects of the evolution of the galaxies.
The star-forming to blue ratio is much higher in the cluster virial regions than in the field, most likely because of the different star formation histories of the galaxies in the different global environments. 

\subsubsection*{\it SFR-Mass relation}
Above the mass completeness limit, at all redshifts and considering both parametrisation of environment, galaxies in the virial/densest regions and galaxies in the field/less dense regions occupy the same locus of the plane, indicating no strong environmental effects at play. Comparing the galaxies at different redshifts, at fixed stellar mass we recover the well-known decline in SFR with time. At any given redshift, the median SFR as a function of mass is similar in all environments. Nonetheless, an important difference emerges between the global and local parametrisations. When using the former, a population of galaxies with reduced SFR compared to the expected value given their stellar mass is detected in the cluster virial regions. These are likely to be in transition from star forming to passive. Their incidence increases going from the higher towards lower redshifts. Such a population is not detected when comparing the SFR-mass relation of galaxies in two extreme bins of LD. 

This dichotomy emerging in the galaxy properties when investigated in either a global or local environment framework are intrinsically related to the different physical meaning of the two parametrisations.
The potential well of X-ray groups and clusters must enhance physical processes related to the presence of the dark matter halo and the hot intra-cluster medium on one side, whereas high-LD regions select associations of galaxies which are physically close and thus more prone to interactions and encounters with other galaxies.

Whether these two definitions insinuate differences in the star formation histories of the involved galaxy populations will be investigated in detail in Guglielmo et al. (in preparation).
In fact, the availability of full spectral fitting results on the galaxy sample explored in this paper enables us to follow a complementary approach, and trace the histories of individual galaxies to examine how the SFH proceeded in X-ray clusters, in the field and in high-/low- local overdensities of galaxies. This technique was already exploited in \cite{Guglielmo2015} in a low-redshift sample of galaxies in clusters and in the field, which can then be used as basis for comparison with the local Universe population.

\begin{acknowledgements}

We acknowledge Lucio Chiappetti for his careful technical report, which guarantees the conformity of all the XXL papers and helped us to improve our work.
XXL is an international project based around an XMM Very Large Programme surveying two 25 deg$^2$ extragalactic fields at a depth of $\sim 5 \times 10^{-15} {\rm erg \, s^{-1} \, cm^{-2}}$ in the [0.5--2] keV band for point-like sources. The XXL website is http://irfu.cea.fr/xxl. Multi-band information and spectroscopic follow-up of the X-ray sources are obtained through a number of survey programmes, summarised at http://xxlmultiwave.pbworks.com/. The Australia Telescope Compact Array is part of the Australia Telescope National Facility, which is funded by the Australian Government for operation as a National Facility managed by CSIRO.
GAMA is a joint European-Australasian project based around a spectroscopic campaign using the Anglo-Australian Telescope. The GAMA input catalogue is based on data taken from the Sloan Digital Sky Survey and the UKIRT Infrared Deep Sky Survey. Complementary imaging of the GAMA regions is being obtained by a number of independent survey programmes including GALEX MIS, VST KiDS, VISTA VIKING, WISE, Herschel-ATLAS, GMRT, and ASKAP providing UV to radio coverage. GAMA is funded by the STFC (UK), the ARC (Australia), the AAO, and the participating institutions. The GAMA website is http://www.gama-survey.org/.
This work was supported by the Programme National Cosmology et Galaxies (PNCG) of CNRS/INSU with INP and IN2P3, co-funded by CEA and CNES.
\end{acknowledgements}

\bibliographystyle{aa}
\bibliography{bibliography.bib}

\begin{appendix}

\section{Spectroscopic completeness}
\label{compl_w1}

In this section we describe in detail the methodology we applied to compute the spectroscopic completeness of our sample.
To properly account for the redshift dependence of the completeness ratio, in addition to considering separately  three redshift bins (0.1$\leq$z$<$0.2, 0.2$\leq$z$<$0.3, 0.3$\leq$z$\leq$0.5), we base our procedure on the combined use of spectroscopic and photometric redshifts.
Specifically, in each redshift bin, the sampling rate (SR) is defined as the ratio of the number of objects with spectroscopic redshift to the number of possible targets (i.e. the photo-z sample). 

The steps taken to compute the completeness can be summarised as follows. 
Considering the galaxies in the spectrophotometric sample with reliable measurements of both spectroscopic and photometric redshift, and defined a redshift range of interest, we call
\begin{itemize}
\item[] - $N_{11}$= the number of objects with spectroscopic redshift in the selected redshift range, and photo-z in the same range.
\item[] - $N_{12}$= the number of objects with spectroscopic redshift in the selected redshift range, but photo-z not in the same range.
\item[] - $N_{21}$= the number of objects with spectroscopic redshift out of the selected redshift range, but photo-z in the range.
\item[] - $N_{22}$= the remaining number of objects with both spectroscopic and photo-z outside the selected redshift range.
\end{itemize}

These numbers are  used to define the two fractions that allow us to compute the expected number of objects relative to the entire photo-z sample, which also includes galaxies with no spectroscopic redshift, starting from the spectrophotometric sample,
\begin{equation}
f_{1}=\frac{N_{11}}{(N_{11}+N_{21})}
\label{f1_eq}
\end{equation}
is the fraction of all objects with photo-z in the selected redshift range that truly belong to the range (i.e. with spectroscopic redshift in the range).
Then,
\begin{equation}
f_{2}=\frac{N_{12}}{(N_{12}+N_{22})}
\label{f2_eq}
\end{equation}
is the fraction of all objects with photo-z outside the range that are instead within the considered redshift bin (i.e. with spectroscopic redshift in the range). These objects should be considered in the SR estimate of the given redshift slice even if their photo-z would not include them.

These two fractions are finally used to estimate the number of expected photo-z objects in the range, when applied to the whole photo-z sample,
\begin{equation}
N_{exp} = f_1 \times N_{photo-z, in} + f_2 \times N_{photo-z, out}
\end{equation}
Where the numbers $N_{photo-z, in}$ and $N_{photo-z, out}$ refer, respectively, to the number of objects with photo-z in and outside the selected redshift range in the total photo-z sample.

The sampling rate is finally defined as
\begin{equation}
SR= \frac{(N_{11}+N_{12})}{N_{exp}}
\end{equation}
where ($N_{11}+N_{12}$) is the total number of galaxies with spectroscopic redshift in the selected redshift range.

By construction, the sum of the inverse of the SRs, i.e. the spectroscopic weights, at all redshifts and in the magnitude limited sample approximately gives the number of objects in the magnitude limited parent photo-z sample; small differences can be due to the different redshift range covered by the spectroscopic and photo-z sample.

To account for the dependence on the different SR of spectroscopic surveys in different regions in the sky, we proceed as already performed in XXL Paper XXII and subdivide the field in three stripes of declination and we further divide each stripe in RA creating a grid of 1.0 deg width. Finally, we consider intervals of 0.5 $r$-band observed magnitude in the 22 resulting cells (see Fig \ref{RA_DEC_sample}).

From these results, we obtain the spectroscopic completeness curves as the SR as a function of magnitude, in all the sky cells and redshift bins in which the sample has been divided.
\begin{figure*}
\begin{center}
\includegraphics[scale=0.39]{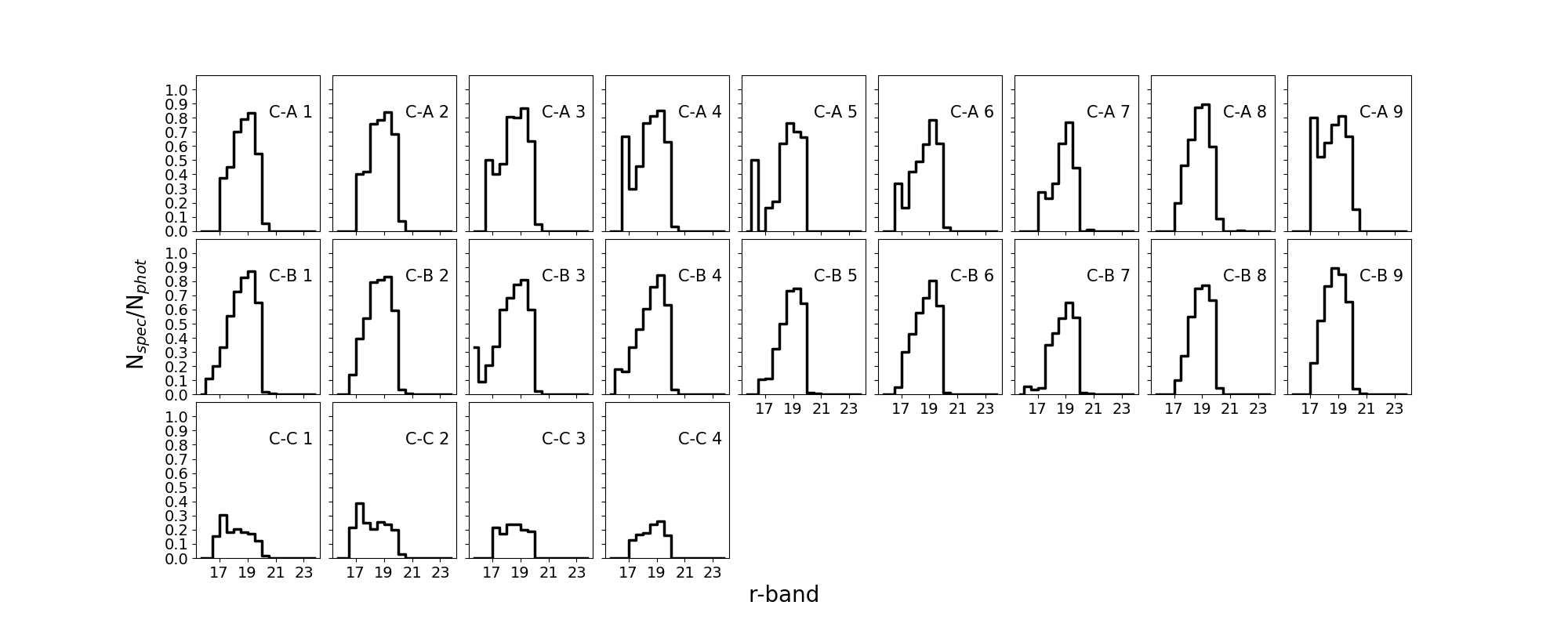}
\includegraphics[scale=0.39]{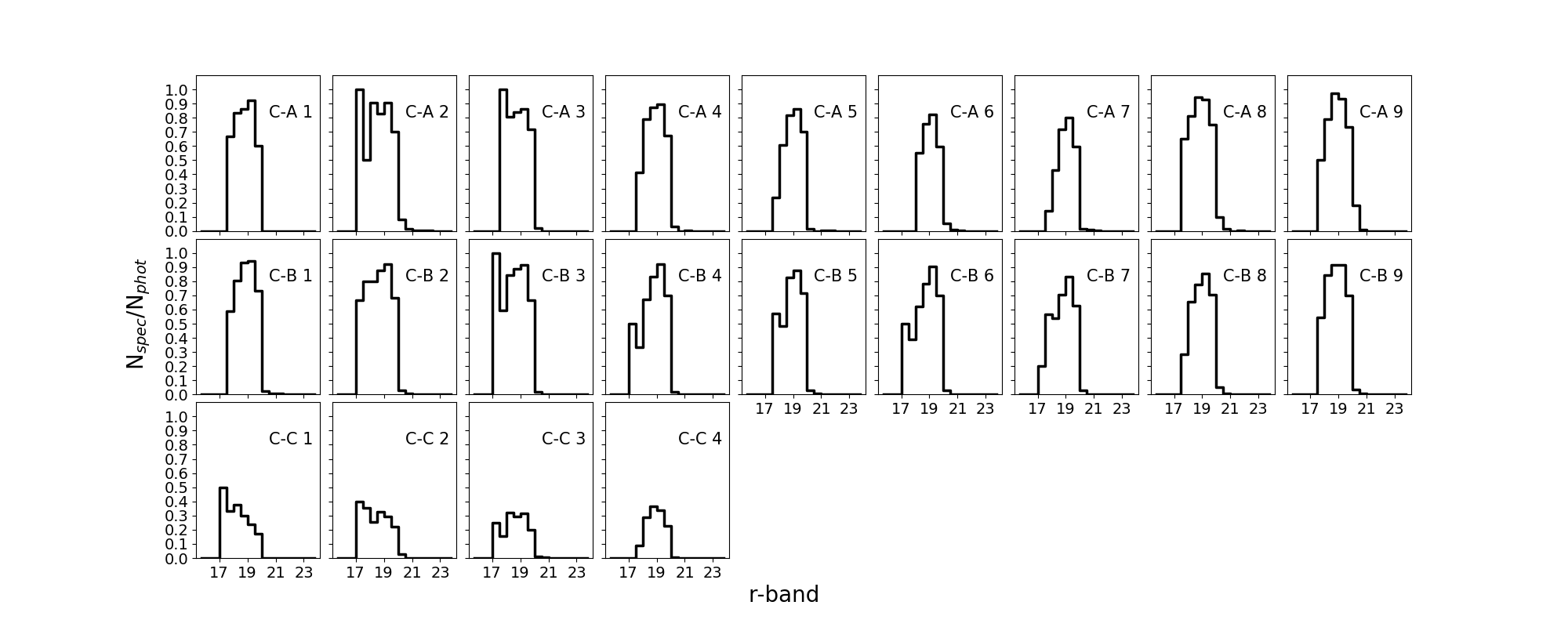}
\includegraphics[scale=0.39]{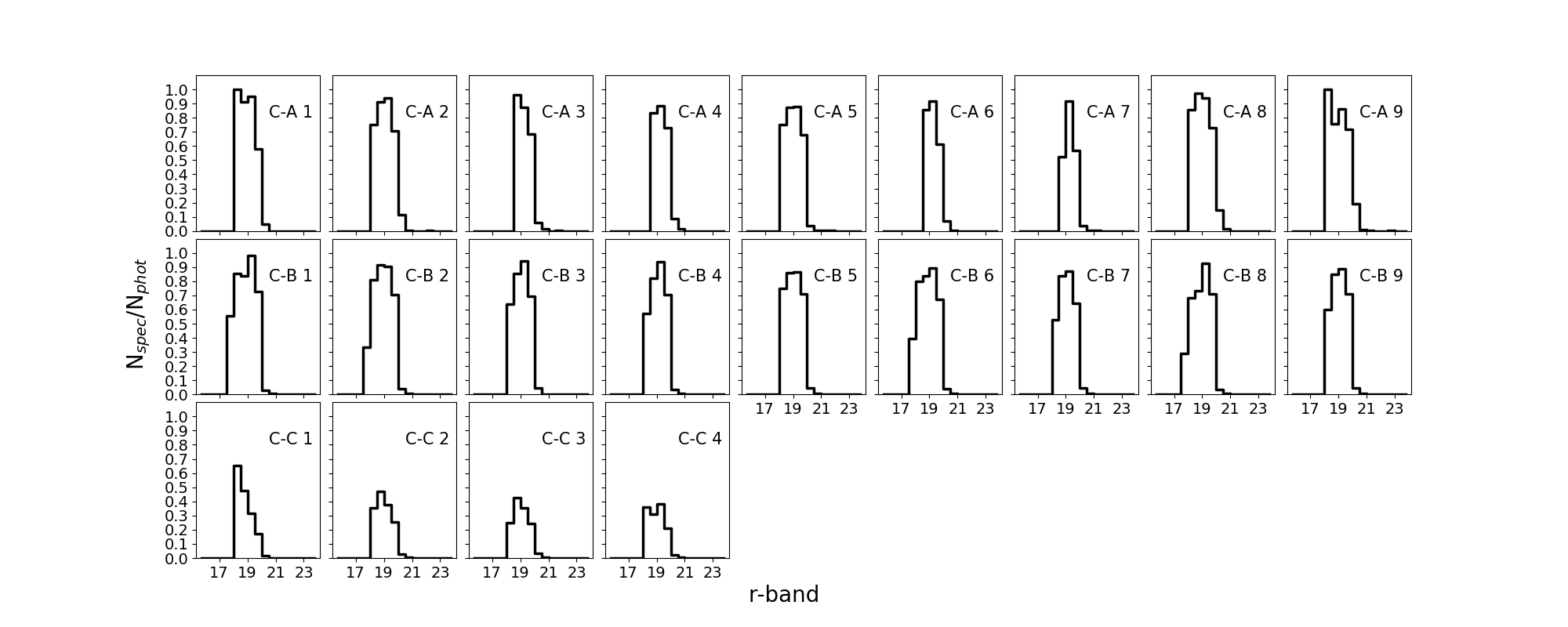}
\caption{Completeness curves computed in three redshift bins and in different RA-DEC cells in the sky, as explained in the main text. From the top to the bottom panel, the represented redshift ranges are respectively $0.1 \leq z < 0.2$, $0.2 \leq z < 0.3$, $0.3 \leq z \leq 0.5$.}
\label{compl_w1_LePh}
\end{center}
\end{figure*}

Such curves are shown in Figures \ref{compl_w1_LePh}.

Each galaxy is finally weighted for the inverse of the SR computed as explained in above, which accounts for its redshift, position in the sky, and observed magnitude.

\section{AGN contamination}\label{app:AGN}
To compute the number of (broad and narrow lines) AGNs in our galaxy sample we consider all galaxies having a GAMA spectrum (98\% of all galaxies in the magnitude complete sample), and perform our classification using the DR3 GAMA catalogues which include the region overlapping with the XXL-N field (GAMA-G02, \citealt{Baldry2018}).

The catalogues contain emission line measurements derived with Gaussian fits of different complexity to the lines as well as specific parameters used to identify peculiar galaxies.
First, we remove duplicates in the spectra and selected the best-observed spectrum with the flag IS\_BEST=True in the catalogues, and we consider only reliable redshift measurements (NQ$\leq$3).
We proceed separately for broad- and narrow-line AGNs as follows:

\begin{itemize}
\item Broad-line AGN: These objects can be simply identified by means of the output of the model selection (HA\_MODSEL\_EMB\_EM in the GaussFitComplex Table), as explained in Sect. 2.3 in \cite{Gordon2017}. Broad-line AGNs are characterised by a value of the model selection score parameter greater than 200, and have a S/N on the broad H$\alpha$ component $>$3 (HA\_B\_FLUX/HA\_B\_FLUX\_ERR$>$3 in the GaussFitComplex Table).

\item Narrow-line AGN: For these objects, we rely on the table containing simple Gaussian fit to the spectral lines (GaussFitSimple), and consider the classification of \cite{Gordon2018} based on line ratios.
We select only spectra with reliable S/N on the interesting lines (NIIR\_FLUX/NIIR\_FLUX\_ERR$>$3 for the NII line, and HA\_FLUX/HA\_FLUX\_ERR$>$3 for the H$\alpha$ line), and correct the H$\alpha$ line for stellar absorption applying Eq 5 from \cite{Gordon2017} (a GAMA specific version of Eq 4 from \citealt{Hopkins2003}) as follows:
\begin{equation}
F_{cor}=\frac{EW+2.5 \AA}{EW} F_{obs}
\end{equation} 

\end{itemize}

where F$_{cor}$ is the corrected flux measurement, F$_{obs}$ is the observed flux measurement (HA\_FLUX in the GaussFitComplex Table), and EW is the measured equivalent width of the emission line (HA\_EW in the GaussFitComplex Table).
Finally, we classify as likely narrow-line AGN those galaxies having log$_{10}$([NII],$\lambda$6583/H$\alpha$)$>$0.4.
We point out that this classification based on two emission lines is the most conservative and may also include normal galaxies with high SFRs, making the AGN contribution evaluated in this work an upper limit.

Having classified and flagged broad- and narrow-line AGN, we crossmatch the catalogue of spectra with our spectrophotometric catalogue and compute their upper limit fraction with respect to the number of star-forming galaxies in the three usual redshift bins and in the magnitude complete sample (similar fractions are also found in the mass limited sample):
\begin{itemize}
\item 0.1$\leq$z$<$0.2: 762/5026=15.2\%
\item 0.2$\leq$z$<$0.3: 1166/5817=20.04\%
\item 0.3$\leq$z$<$0.5: 372/3047=12.2\%
\end{itemize}

\section{Local density}
\label{LD_computation}

We compute the LD of galaxies in the spectrophotometric sample taking as a reference the photo-z sample used in the spectroscopic completeness computation, and considering one redshift bin at a time.
The LD around each galaxy is given as the ratio of the number of galaxies in the parent photometric-redshift sample per unit of projected {\it comoving} area on the sky.
Our method proceeds through the following different phases:
\begin{itemize}
\item[] - Computation of the observed magnitude limit used to select galaxies in the sample as a function of redshift. To perform the same sample selection, we apply the same absolute magnitude cut in all the redshift slices. The value is selected in order to balance the error in the photo-z estimate, which increases towards fainter magnitudes, and the propagation of the observed magnitude down to redshift 0.1, and thus to minimise the loss of galaxies occurring with brighter observed magnitude cuts.
We consider as observed magnitude limit $r$=23.0 at z=0.5 and compute the corresponding absolute magnitude as follows: 
\begin{equation}
M_r = r - 5 \cdot (log_{10} D_L -1) - K_{corr}.
\end{equation}
where $r$ is the observed r-band magnitude and $D_L$ is the luminosity distance in pc. The value $K_{corr}$ is the K-correction that takes into account that the same photometric filter samples different spectral ranges when applied to the SED of galaxies at different redshifts and is taken from \cite{Poggianti1997}, assuming the typical value of an intermediate type galaxy (Sab) in r band at the selected redshift. The application of this formula leads to an absolute magnitude of M$_r$ = -19.89, which is then converted into an observed magnitude limit as a function of redshift by means of the inverse formula,
\begin{equation}
r(z) = -19.89 + 5 \cdot (log_{10} D_L(z) -1) + K_{corr}(z) + P.E.(z)
\label{r_z}
\end{equation}
where the $D_L$ is computed at the redshift of the considered galaxy, $K_{corr}$ is a function of redshift, and P.E.(z) is the passive evolution of galaxies, which becomes redder with decreasing redshift as a consequence of the ageing of their stellar population; the correction for passive evolution is 0.1 mag each $\Delta z$=0.1 \citep{Poggianti2008}.

\item[] - Computation of the number of galaxies in the spectrophotometric sample within a {\it comoving} circle of 1 Mpc radius at the redshift of the galaxy in the centre and within a redshift range of $\pm 0.05$ with respect to the redshift of the same galaxy. To account for uncertainties in the photo-z measurements, we estimate the expected number of galaxies in the photo-z sample in the considered redshift range around the selected galaxy with the same method used for the spectroscopic completeness. We define the fractions $f_1$ and $f_2$ given in equations \ref{f1_eq} and \ref{f2_eq} in the spectrophotometric sample and use them to weight the photo-z sample and compute $N_{exp}$. This value represents the correct number counts within the {\it comoving} projected area of 1 Mpc radius around the galaxy. The area of the circle is then computed and the LD is defined as the ratio of the two quantities.

\begin{figure*}
\begin{center}
\includegraphics[scale=0.65]{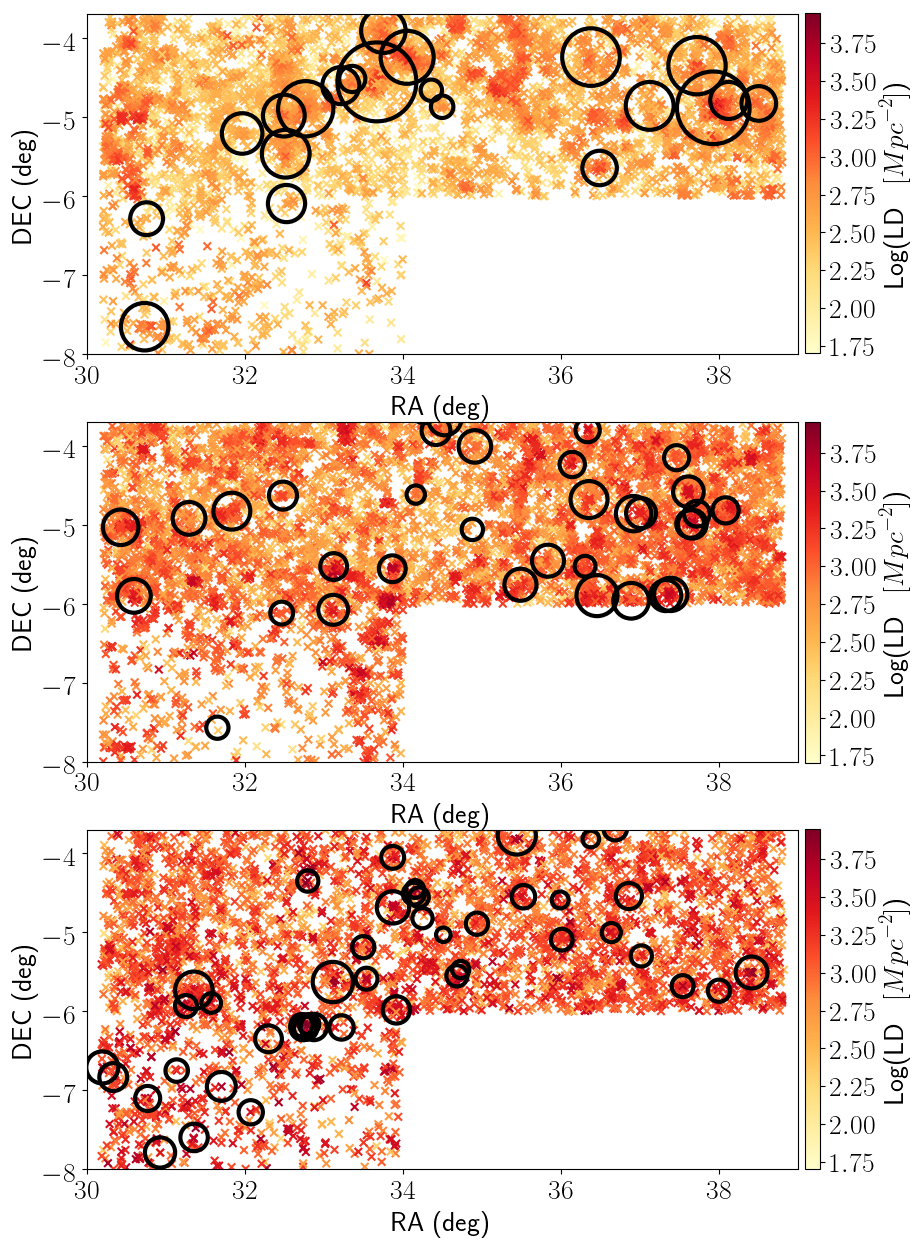}
\caption{Spatial distribution in the sky of the spectrophotometric magnitude limited sample. Data points are colour coded according to their log(LD), after a sigma-clipping has been performed on the parent distribution. From the top to bottom panel the represented redshift bins are $0.1 \leq z < 0.2$, $0.2 \leq z < 0.3$, $0.3 \leq z \leq 0.5$, respectively. Each panel contains the 3r$_{200}$ extensions of the clusters at the redshift of the bin, represented with black empty circles.
\label{RA_DEC_ld_w1}}
\end{center}
\end{figure*}

\item[] - Correction for edge effects in the field. For galaxies located at the edges of the XXL-N field, we correct the circular area for the fraction of area effectively covered by the data points, and therefore remove empty circular sectors. 
We adopt a numerical solution based on a Monte Carlo simulation method. We generate a circular homogeneous distribution of data points by populating a circle with a sufficiently high number of points (100000) and  compute the zone of exclusion with respect to the edge conditions of the field as the ratio of the number of points falling outside the edges to the total number of points included in the circle. The area of the circle in physical units 
that has to be considered in the LD calculation is then the total {\it comoving} area multiplied by the fraction of area included in the field, $\rm f_{in}=1-f_{out}$, where $\rm f_{out}$ is the fraction of area falling outside the galaxy field.

\end{itemize}

The LD is finally expressed as the logarithm of the quantity computed in the procedure outlined above, with dimension [LD]=$\rm Mpc^{-2}$.

Figure \ref{RA_DEC_ld_w1} reports the spatial distribution of galaxies in the spectrophotometric sample colour coded for the   LD measures.
Each panel also reports the circle of 3$r_{200}$ radius of the clusters in each redshift bin; as expected, in most of the cases, galaxies within the circles are characterised by high LD values.
\end{appendix}

\end{document}